\documentclass[fleqn,usenatbib]{mnras}

\usepackage[T1]{fontenc}

\DeclareRobustCommand{\VAN}[3]{#2}
\let\VANthebibliography\thebibliography
\def\thebibliography{\DeclareRobustCommand{\VAN}[3]{##3}\VANthebibliography}

\usepackage{graphicx}	% Including figure files
\usepackage{amsmath}	% Advanced maths commands
\usepackage{amssymb}	% Extra maths symbols
\title[Radiative feedback on metal-poor stellar IMF]
{Impact of radiative feedback on the initial mass function of metal-poor stars}

\author[S. Chon, T. Hosokawa, K. Omukai, \& R. Schneider]{
Sunmyon Chon $^{1,2}$\thanks{E-mail: sunmyon@MPA-Garching.MPG.DE},
Takashi Hosokawa $^{3}$,
Kazuyuki Omukai $^{2}$, and
Raffaella Schneider $^{4,5,6}$
\\
$^{1}$Max-Planck-Institut f$\ddot{u}$r Astrophysik, Karl-Schwarzschild-Str. 1, D-85741 Garching, Germany\\ 
$^{2}$Astronomical Institute, Graduate School of Science, Tohoku University, Aoba, Sendai 980-8578, Japan\\
$^{3}$Department of Physics, Kyoto University, Kyoto 606-8502, Japan\\
$^{4}$Dipartimento di Fisica, Universit\`{a} di Roma ‘La Sapienza’, P.le Aldo Moro 2, I-00185 Roma, Italy \\
$^{5}$INAF/Osservatorio Astronomico di Roma, via di Frascati 33, I-00078 Monteporzio Catone, Italy \\
$^{6}$INFN, Sezione di Roma 1, P.le Aldo Moro 2, I-00185 Roma, Italy
}

\date{Accepted XXX. Received YYY; in original form ZZZ}

\pubyear{2023}

\begin{document}
\label{firstpage}
\pagerange{\pageref{firstpage}--\pageref{lastpage}}
\maketitle

\begin{abstract}
The stellar initial mass function (IMF) in the early universe is essential to understand the formation of ancient galaxies. To this end, we conduct a series of long-term radiation hydrodynamic simulations following star cluster formation, varying the metallicity from $Z/Z_\odot = 10^{-4}$ to $1$. We particularly consider the effects of protostellar radiative feedback, which modify the exact shape of the IMF and determine the star formation efficiency (SFE), i.e. the ratio between the mass in stars and the initial gas mass in the parental cloud. Our results show that the IMF changes from a Salpeter-type to a top-heavy function as the metallicity decreases. When $Z/Z_\odot \lesssim 10^{-2}$, the IMF becomes log-flat and distinct from a Salpeter-like IMF. Stellar feedback is effective in shaping both the low- and high-mass ends of the IMF. Heating of dust grains by stellar radiation suppresses small-scale fragmentation and reduces the number of low-mass stars with $M_* \lesssim 1~M_\odot$ at all metallicities. The ionizing radiation hinders the growth of massive stars, steepening the slope of the IMF at the high-mass end. The resulting feedback is more effective at lower metallicity, and star formation is regulated by stellar radiative feedback, with the SFE decreasing with decreasing metallicity. We suggest that the unexpectedly large number of UV-bright galaxies at $z>10$ reported by JWST observations can be explained by considering star cluster formation at $Z/Z_\odot \sim 10^{-2}$ or $10^{-3}$, where the IMF is top-heavy, but the SFE is not too low due to stellar feedback.
\end{abstract}

%%%%%%%%%%%%%%%%%%%%%%%%%%%%%%%%%%%%%%%%%%%%%%%%%%

%%%%%%%%%%%%%%%%% BODY OF PAPER %%%%%%%%%%%%%%%%%%

\section{Introduction}
The stellar initial mass function (IMF) is fundamental to understanding the formation and evolution of galaxies. It determines the proportion of massive stars that are formed, whose feedback has a major impact on the environment. Their intense ultraviolet radiation heats and ionizes the interstellar and intergalactic medium (ISM/IGM), eventually leading to cosmic reionization. Stars with $M_* \gtrsim 8~M_\odot$ end their lives in energetic events such as supernovae (SNe) and gamma-ray bursts \citep[e.g.][]{HegerWoosley2002,Fryer+2022}, dispersing synthesized heavy elements into the ISM/IGM \citep[e.g.][]{Ritter+2015, Chiaki+2018, Magg+2018}. More massive stars can collapse directly into black holes (BHs), creating binary BHs that merge and emit gravitational waves \citep{Kinugawa+2014, Schneider+2017, Graziani+2020,Tanikawa+2022, LIGOO3+2023}, and sowing the seeds of supermassive BHs \citep{Bromm&Loeb2003, Latif+2013, Valiante+2016, Chon+2016, Chon+2018, Sessano+2021}.

Hydrodynamical simulations indicate that the IMF in the early universe was substantially different from that in the present-day universe. In pristine environments, the typical stellar mass is estimated to be several $10 - 100~M_\odot$
 \citep{Hirano+2014, Hirano+2015, Susa+2014, Hosokawa+2016, Stacy+2016, Sugimura+2020}, while some lower-mass stars may also form \citep{Machida+2008b, Clark+2011, Machida&Doi2013, Greif+2012, Stacy+2016, Latif+2022}. On the contrary, the typical stellar mass in the present-day universe is in the range of $0.1 - 1~M_\odot$ \citep[e.g.][]{Salpeter1955, Kroupa2002, Chabrier2003}. On the other hand, observations in the solar neighborhood suggest that the IMF has a universal shape with minimal variations in different environments \citep{Bastian+2010}.

The thermal evolution of a pre-stellar cloud is a key element in understanding the variations of the IMF. The effective adiabatic index $\gamma_\text{eff} \equiv \mathrm{d} \log P / \mathrm{d} \log \rho$, where $P$ and $\rho$ are the gas pressure and density, determines the fragmentation mass scale \citep{Larson1985, Bonnell+2004}. When $\gamma_\text{eff}$ is less than one, the temperature decreases as the cloud contracts, causing the cloud to become filamentary in shape \citep{Inutsuka&Miyama1992, Li+2003, Jappsen+2005}. When cooling becomes inefficient and $\gamma_\text{eff}$ becomes greater than one, the filamentary collapse stops and the cloud fragments into spherical cores, thus establishing the scale of fragmentation mass. This characteristic mass scale should also be imprinted on the stellar IMF.

The thermal evolution of a cloud is significantly influenced by its metallicity. Metals can increase the cooling capability through metal line emission and dust thermal emission. A higher metallicity in the cloud leads to a lower temperature and Jeans mass, which could result in a decrease in the typical stellar mass \citep{Omukai2000, Bromm+2001, Schneider+2003, Schneider+2006, Schneider+2012, Omukai+2005, Schneider&Omukai2010, Chiaki+2014}. 
Recent numerical simulations have demonstrated that a finite amount of metals cause the formation of low-mass fragments/stars of $0.01$--$1~M_\odot$ \citep{Tsuribe&Omukai2006, Tsuribe&Omukai2008, Clark+2008, Dopcke+2011, Dopcke+2013, Safranek-Shrader+2016, Chiaki+2016, Chiaki+2021, Shima&Hosokawa2021}. 

In \citet{Chon+2021b} (Paper~I), we conducted hydrodynamic simulations to explore the formation of star clusters in clouds with different metallicities. We resolved down to a spatial scale of approximately 1~au and a mass scale of 0.01~$M_\odot$, and found that the IMF is dependent on the metallicity. In a primordial, metal-free environment, the mass spectrum is approximated by a top-heavy log-flat shape. As the metallicity increases, a low-mass Salpeter-like component emerges and grows in proportion. Consequently, the mass spectrum is composed of the two components, with an increasing fraction of the low-mass component. Above a threshold metallicity of around 0.1~$Z_\odot$, the dust cooling induces fragmentation, resulting in many low-mass stars, so that the mass spectrum is purely composed of the Salpeter-like component.

%%%%%

The thermal evolution of clouds is also altered by stellar radiative feedback effects, which were not included in \citet{Chon+2021b}:  i) the heating of dust grains, and ii) the ionization and heating of gas by the UV radiation. The former tends to increase the peak stellar mass around the low-mass end by raising the local Jeans mass \citep[e.g.][]{Bate2009, Krumholz+2012}. Numerical simulations demonstrate that dust heating suppresses fragmentation also at low metallicities with $Z \sim 10^{-2}~Z_\odot$ \citep[e.g.][]{Safranek-Shrader+2016, Matsukoba+2022}. \citet{Chon+2022} (Paper~II) found that elevated cosmic microwave background (CMB) radiation works similarly to suppress the formation of low-mass stars.  
On the other hand, gas ionization and heating by UV radiation regulate the IMF at the high-mass end by inhibiting mass accretion onto massive stars. Radiation hydrodynamics (RHD) simulations demonstrate that the ionizing feedback determines the final stellar masses when isolated star formation occurs in metal-free or very metal-poor environments \citep{Hosokawa+2011, Fukushima+2020a}. In the case of clustered star formation, the IMF slope at the high-mass end becomes steeper due to the ionizing feedback \citep{He+2019}.

%%%%%%%

Ionization feedback caused by the expansion of HII regions quenches the cloud-scale star formation and also determines the star formation efficiency (SFE), i.e., the conversion ratio from the gas to the stars. In metal-enriched environments, especially where $Z/Z_\odot \gtrsim 10^{-2}$, clouds fragment into numerous stars, leading to the formation of star clusters that span the entire cloud scale. 
Star formation in these environments is suppressed by ionizing photons from a group of stars. 
Numerical simulations have shown that the SFE decreases with decreasing metallicity \citep{He+2019, Fukushima+2020b, Guszejnov+2022}.
In lower-metallicity environments with $Z/Z_\odot \lesssim 10^{-3}$, a single or a few massive stars form inside a parental cloud instead of a star cluster. 
These stars have masses ranging from several tens to hundreds of solar masses, and star formation is mainly quenched by feedback from the central massive star \citep{Hirano+2015, Sugimura+2020, Fukushima+2020a}. However, it is still unclear how the SFE varies from present-day to very metal-poor environments, where the stellar distribution is notably different.

In this paper, we investigate the characteristics of stellar clusters that form at different metallicities, by performing a suite of RHD simulations. In particular, we study the long-term evolution until the radiative feedback from forming protostars destroys the original cloud, allowing us to systematically study the metallicity dependence of both the IMF shape and SFE.
Recent JWST observations have revealed the existence of a large number of UV-bright galaxies at $z \gtrsim 10$ that exceeds the predictions of standard star and galaxy formation models at very high redshifts \citep[e.g.][]{Harikane+2023}. 
One possible solution is non-standard star formation with some combination of top-heavy IMFs and high SFEs \citep{Inayoshi+2022,Trinca+2023}, both of which are provided by our simulations as functions of metallicity.

The paper is structured as follows. We describe the initial condition and the numerical methodology in Section~\ref{sec::method}.
We present our numerical results in Section~\ref{sec::results} and discuss the implication of our results in Section~\ref{sec::discussion}.
We summarize our findings in Section~\ref{sec::summary}.

\section{Methodology} \label{sec::method}
We use the {\tt GADGET-2} code \citep{Springel2005} to simulate the formation of star clusters in clouds of varying metallicities, from very low to present-day levels, [Z/H] $\equiv \log Z/Z_\odot = -4$ -- $0$. To account for the long-term effects of stellar radiative feedback, we have added several radiative processes to the original {\tt GADGET-2} code. In this section, we provide a brief overview of the numerical methodology, which is mostly the same as in our previous studies (Paper I and II), and describe the implementation of the radiative process in our code (Section~\ref{sec::radiative_process}).

\subsection{Simulation setup}
We begin each simulation run with a critical Bonnor-Ebert sphere of a central gas density of $10^3~\mathrm{cm^{-3}}$ and a temperature of $200~$K, but with density enhancement by a factor of $1.2$ to induce gravitational collapse. 
This setup has a mass of $M_\text{cloud} = 6300~M_\odot$ and a radius of $R_\text{cloud} = 1.2 \times 10^6~\mathrm{au} = 5.9~\mathrm{pc}$.
The initial cloud is rigidly rotating with $\Omega = 2.08\times 10^{-15} \mathrm{s^{-1}}$.
The ratio of rotational energy to gravitational energy is $0.001$, comparable to the observed values of Galactic molecular cloud cores \citep[$10^{-4}$ -- $0.07$,][]{Caselli+2002}. 
We further impose a turbulent velocity field with a trans-sonic Mach number of $1$, for which the mass-weighted velocity dispersion is equal to the sound speed. 
This value also represents the turbulent velocity observed for Galactic molecular clouds at scales of $\sim~$pc \citep{Larson1985}. We generate a random turbulent velocity field following the power spectrum $P(k) \propto k^{-2}$ on $128^3$ grids \citep[e.g.][]{MacLow1999}. We assign the turbulent velocity to each gas particle by interpolating from grids that are close by.

The initial particle mass is $4.7\times 10^{-3}~M_\odot$. To follow the cloud collapse across a wide dynamic range of densities, we employ a two-level particle splitting technique when the gas density exceeds certain thresholds $n_\text{th} = 10^5$ and $10^8~\mathrm{cm^{-3}}$. Following the approach outlined in \citet{Kitsionas+2002}, we split a gas particle into $13$ daughter particles of equal mass. 
The masses of the daughter particles in the first and second levels of splitting are $3.62 \times 10^{-4}$ and $2.79 \times 10^{-5}~M_\odot$, respectively.  
The minimum resolvable mass is then $M_\text{res} \sim 1.5 N_\text{neighb} M_\text{part} = 2.68 \times 10^{-3}~M_\odot$, where $N_\text{neighb}=64$ is the number of SPH particles within the kernel and $M_\text{part}$ is the mass of a single SPH particle \citep{Bate+1995}.
This splitting scheme allows us to resolve the Jeans mass at all gas densities for all metallicity models with the minimum Jeans mass being $\sim 0.01~M_\odot$ \citep{Bate2009, Bate2019}. 

When the gas density exceeds $2\times 10^{15} \mathrm{cm^{-3}}$, a sink particle is inserted with a radius of $r_\text{sink}$ set to twice the smoothing length of the original gas particle, which is approximately $1~$au. To prevent the formation of spurious sink particles, an adiabatic equation of state with $\gamma = 5/3$ is applied for gas densities higher than $10^{15}~\mathrm{cm^{-3}}$ \citep{Chon+2018, Susa2019}. Additionally, the gas particle must be located at a local minimum in the potential field \citep{Hubber+2013} to ensure that the sink particles are placed at the centre of gravitationally bound cores. If the distance between two sink particles is less than the sum of their radii, they are allowed to merge.

\subsection{Chemistry and radiation feedback from the protostars} \label{sec::radiative_process}

We solve the non-equilibrium chemical network of eight primordial species, e$^-$, H, H$^+$, H$^-$, H$_2$, D, D$^+$, and, HD, with 22 chemical reactions among them and consider the thermal processes associated with the primordial species and heavy elements \citep{Yoshida+2006, Matsukoba+2019, Chon+2021a}.

We also consider chemical and thermal processes related to the radiation from protostars, such as photo-ionization of H, dissociation of H$_2$, photo-detachment of H$^-$, and heating of dust grains. We assume that the direct radiation from stars is responsible for photo-ionization and dissociation. To evaluate the optical depth of ionizing UV ($h\nu > 13.6$~eV) and Lyman-Werner (LW) radiation ($11.2 < h\nu < 13.6~$eV) from each radiating star, we use the ray-tracing method, RSPH \citep{Susa2006}. We take into account the four main continuum opacities in the LW band; HI and H$^{-}$ free-bound absorption, Rayleigh scattering by HI, and Thomson scattering. The opacity values for these processes are taken from \citet{Lenzuni+1991} and are evaluated at $h\nu = 11.2~$eV. Additionally, we consider the scattering and absorption by dust grains \citep{Semenov+2003}. We have tested this scheme and confirmed that it successfully captures the expansion of HII regions \citep{Chon+2017, Chon+2021b}. We also calculate the self-shielding effect against H$_2$ dissociation in a similar way to the calculation of $\tau_\text{UV}$.
The shielding factor $f_\text{shield}$ is given by \citep{Draine+1996},
\begin{align}
f_\text{shield} &= 
\min \left \{ 1, \left ( \frac{N_{\text{H}_2}}{10^{-14}~\mathrm{cm^{-2}}} \right )^{-3/4} \right \},
\end{align}
where $N_{\text{H}_2}$ is the column density of H$_2$. The dissociation rate of H$_2$ is given by
\begin{align}
k_{\text{diss}, \text{H}_2} &= 1.39 \times 10^9 (\text{in cgs units}) \sum_i J_{\text{LW}, i} f_{\text{shield}, i} e^{-\tau_{\text{LW},i}}, 
\end{align}
with
\begin{align}
J_{\text{LW}, i} &=  \frac{L_{*,i}}{16\pi \sigma_\text{SB} T_{\text{eff},i}^4 r_i^2} B_{\nu_\text{LW}, i}, 
\end{align}
and
\begin{align}
B_{\nu, i} &= \frac{2h \nu^3}{c^2} \frac{1}{\exp \left ( \frac{h \nu}{k_\text{B} T_{\text{eff},i}} \right ) - 1},
\end{align}
where the subscript $i$ runs over all the radiation sources. Here, $r_i$ represents the distance from the light source $i$, $L_{*,i}$ and $T_{\text{eff},i}$ are the stellar luminosity and effective temperature of source $i$, $h\nu_\text{LW} = 12.4~$eV is the representative energy in the LW band, $\sigma_\text{SB}$ is the Steffan-Boltzmann constant, and $B_\nu$ is the intensity of the Planck distribution at frequency $\nu$.
For the photo-detachment of H$^{-}$, we assume that it is optically thin and the rate is given by,
\begin{align}
k_{\text{pd, H}^{-}} &= \int \frac{J_\nu}{h\nu} \sigma_{\nu, \text{pd}} d\nu, \\
J_\nu &= \sum_i  \frac{L_{*,i}}{16\pi \sigma_\text{ST} T_{\text{eff},i}^4 r_i^2} B_\nu,
\end{align}
where $\sigma_{\nu,\text{pd}}$ is the cross-section of the photo-detachment of H$^-$ provided by \citet{John1988}.

The temperature of dust grains, $T_\text{dust}$, is determined by the equilibrium between the energy absorbed from stellar and cosmic microwave background (CMB) radiation, and the energy lost through collisions with gas and thermal emission from the dust itself:
\begin{align}
4 \sigma_\text{SB} T_\text{dust}^4 \kappa_\text{gr} \rho &= \Lambda_\mathrm{gas\rightarrow dust} + 4 \sigma_\text{SB} T_\text{rad}^4 \kappa_\text{gr} \rho + 4 \sigma_\text{SB} T_\text{CMB}^4 \kappa_\text{gr} \rho, 
\label{eq::Tdust} 
\end{align}
where $\Lambda_\mathrm{gas\rightarrow dust}$ is the rate of energy transfer from the gas to the dust grains through collisions \citep{Tielens+1985}, $T_\text{rad}$ the stellar radiation temperature, and $\kappa_\text{gr}$ the grain opacity.
In this paper, we set the CMB temperature $T_\text{CMB}=2.76~$K assuming $z=0$, meaning that the CMB radiation has a negligible contribution to dust heating. 
This allows us to focus on the metallicity effects, 
while the effects of CMB can be important at the high-redshift universe as we will discuss in Section~\ref{sec::JWST}.
In deriving equation~\eqref{eq::Tdust}, we evaluated the opacities for the stellar and CMB radiation following $\kappa_\text{gr} = \kappa_\text{gr} (T_\text{dust})$ in the right-hand side. 
We approximate the radiation temperature as 
\begin{equation}
T_\text{rad}^4 = \sum_i \frac{L_{*,i}}{16\pi \sigma_\text{ST} r_i^2},
\end{equation}
where we have assumed that the stellar radiation flux is attenuated only geometrically. 
In the steady state with spherical symmetry, this assumption provides a minimum energy density for the stellar radiation and, consequently, for the heating rate. We have found that this prescription is consistent with some previous RHD simulation results, such as the radial dust temperature profiles at $r\lesssim 10^3~$au around accreting protostars \citep{Yorke+1999, Kuiper+2018, Fukushima+2020a}. Note that these simulations solve radiative transfer in the optically-thick diffusive regime. We have implemented the photo-electric heating following \citet{Bakes&Tielens1994}, where the heating rate is determined by a function of the Habing parameter $G_0$ \citep{Habing1968} defined as,
\begin{align}
G_{0} = \frac{4 \pi}{1.6\times 10^{-3}} \sum_i e^{-\tau_{\text{LW},i}} \int J_{\nu,i} d\nu,
\end{align}
and the integration runs for $5.12~\mathrm{eV} < h\nu < 13.6~\mathrm{eV}$.

\begin{figure*}
\includegraphics[width=1.\textwidth]{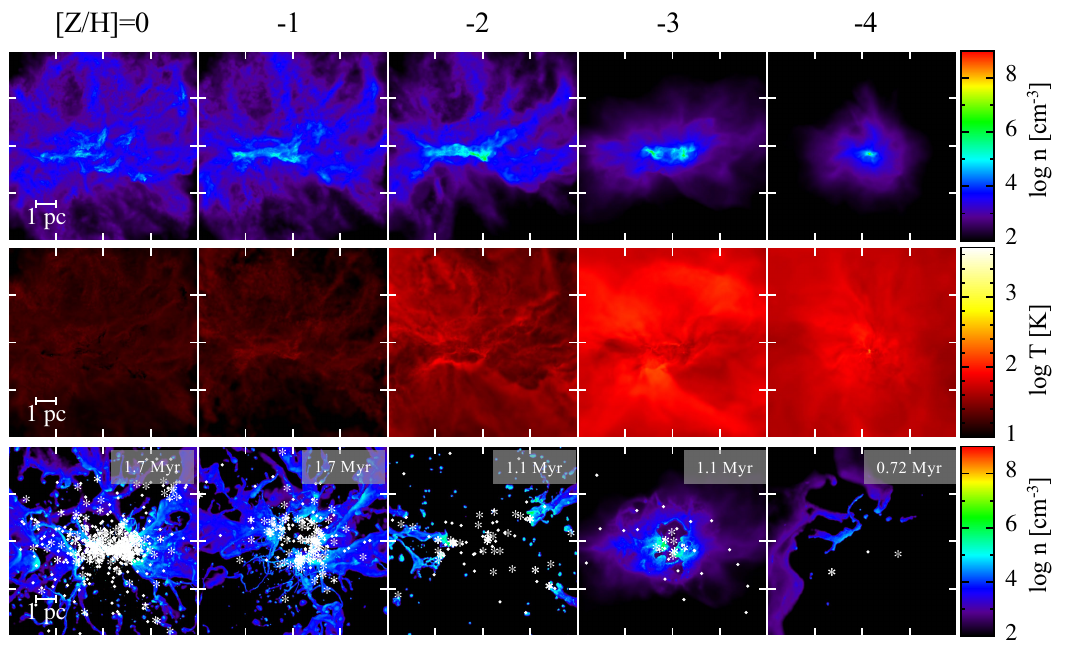}
\caption{
The projected density (top) and temperature distribution (middle rows) at the epoch of the first protostar formation for [Z/H]$=0$, $-1$, $-2$, $-3$, and $-4$. 
The panels in the bottom row show the density distribution at the epoch when the HII region expands around the star cluster and the cloud starts to be disrupted.
We over-plot the stars with a mass larger than (smaller than) $1~M_\odot$ by white asterisks (dots, respectively). We also attach the time since the first protostar is created.
}
\label{fig::snapshots_coll}
\end{figure*}
We model the properties of each star particle, such as internal luminosity $L_*$ and radius $R_*$, as functions of the stellar mass $M_*$ and accretion rate $\dot{M_*}$, using the results of one-dimensional stellar evolution calculations.
To do this, we constructed tables of $L_*$ and $R_*$ as functions of the stellar mass for constant accretion rates of $10^{-5}$, $10^{-4}$, $10^{-3}$, $10^{-2}$, and $10^{-1}~M_\odot\mathrm{yr}^{-1}$ at each metallicity \citep{Hosokawa+2009}. 
The total stellar luminosity $L_\text{tot}$ is the sum of the internal ($L_*$) and the accretion ($L_\text{acc}$) luminosities, which is given by 
\begin{align}
L_\text{tot} = L_* + L_\text{acc} = L_* + f_\text{acc}\frac{G M_* \dot{M_*}}{R_*},
\end{align}
where $f_\text{acc}$ is a non-dimensional parameter set to $0.75$ \citep{Offner+2009}.

The sources of UV radiation are stars with $>5~M_\odot$, undergoing the Kelvin-Helmholtz (KH) contraction during the accretion and eventually reaching the main-sequence stage. Those stars emit a considerable number of UV photons \citep{Hosokawa+2009}, except for cases with very rapid accretion at $\dot{M_*} \gtrsim 10^{-2}~M_\odot\mathrm{yr}^{-1}$ \citep{Hosokawa+2012}.

\section{Result} \label{sec::results}

\begin{figure*}
\includegraphics[width=1.\textwidth]{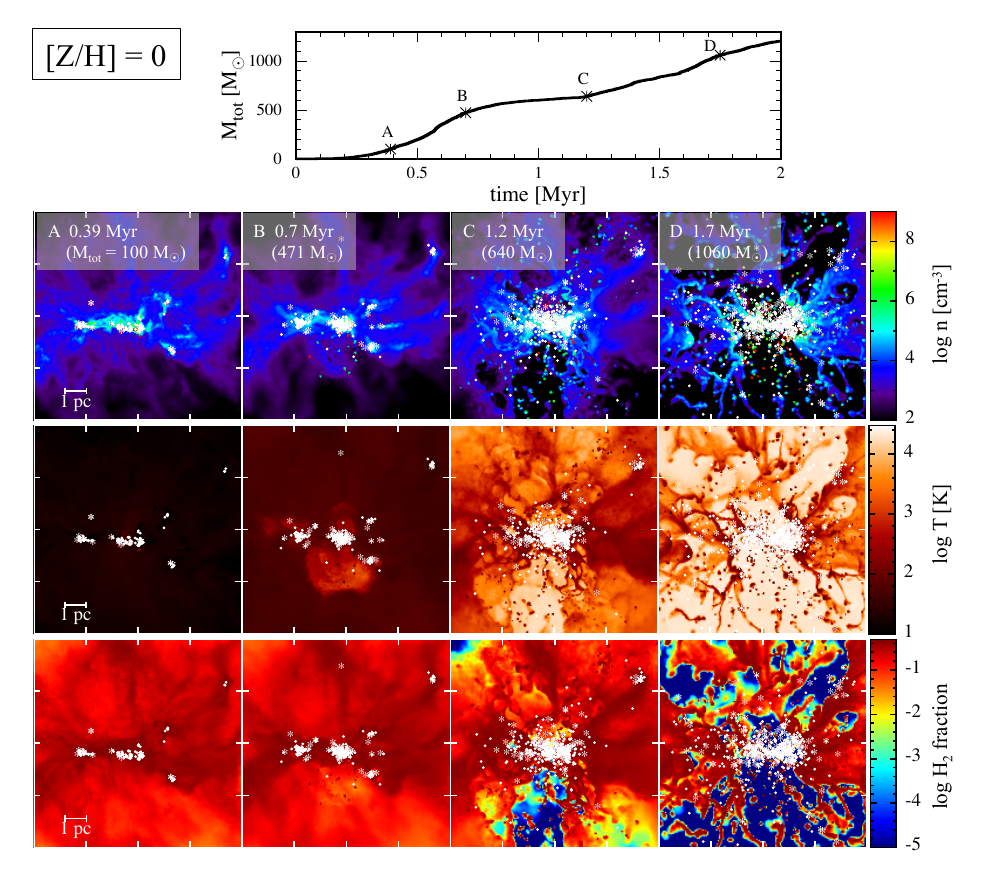}
\caption{
The projected distributions of the density (top), temperature (middle), and the H$_2$ fraction (bottom columns) for the case with [Z/H] $=0$. 
The panels in the different columns depict distributions at different periods: the initial star-forming phase (referred to as epoch A), the time when the HII region begins to expand (epoch B), the point when the HII region is fully extended (epoch C), and the phase when star formation is quenching (epoch D).
We attach the elapsed time since the first protostar formation and the total stellar mass at that epoch. We also superpose the positions of protostars with a mass larger (smaller) than $1~M_\odot$ by the white asterisks (dots, respectively).
}
\label{fig::snapshots_m0}
\end{figure*}

\begin{figure*}
\includegraphics[width=1.\textwidth]{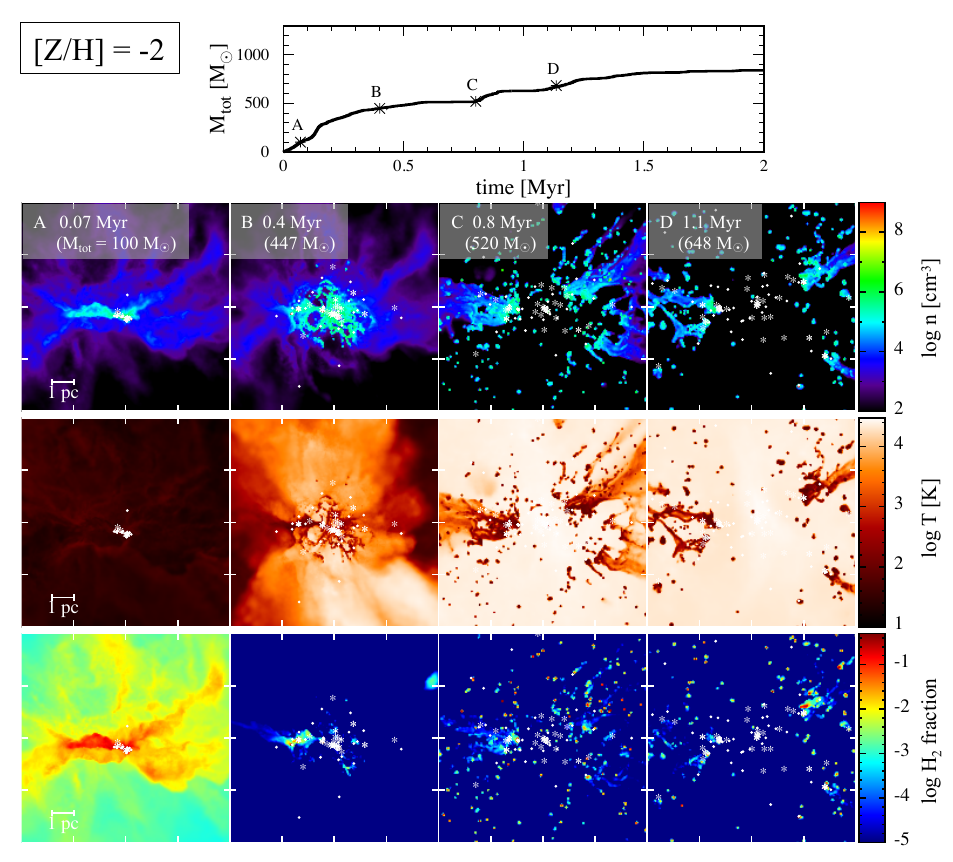}
\caption{
Same as Fig.~\ref{fig::snapshots_m0} but for the case with [Z/H] $=-2$.
}
\label{fig::snapshots_m-2}
\end{figure*}

\begin{figure*}
\includegraphics[width=1.\textwidth]{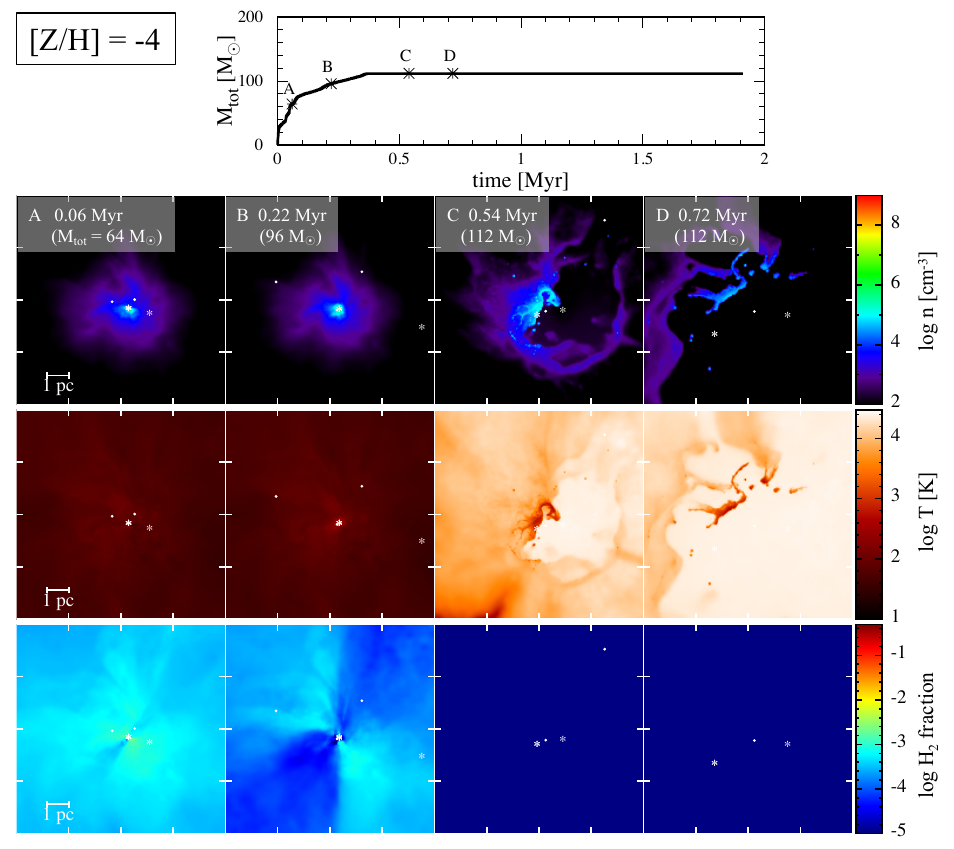}
\caption{
Same as Fig.~\ref{fig::snapshots_m0} but for the case with [Z/H] $=-4$.
Panel B corresponds to the distribution at the epoch when the star formation rate decreases due to the stellar feedback but the HII region does not develop at this moment.
At $t=0.54~$Myr (epoch C), the HII region begins to expand around the ejected runaway massive star.
The star-forming gas is completely evaporated and the star formation is quenched at $t=0.72~$Myr (epoch D).
}
\label{fig::snapshots_m-4}
\end{figure*}

\begin{figure*}
\includegraphics[width=1.\textwidth]{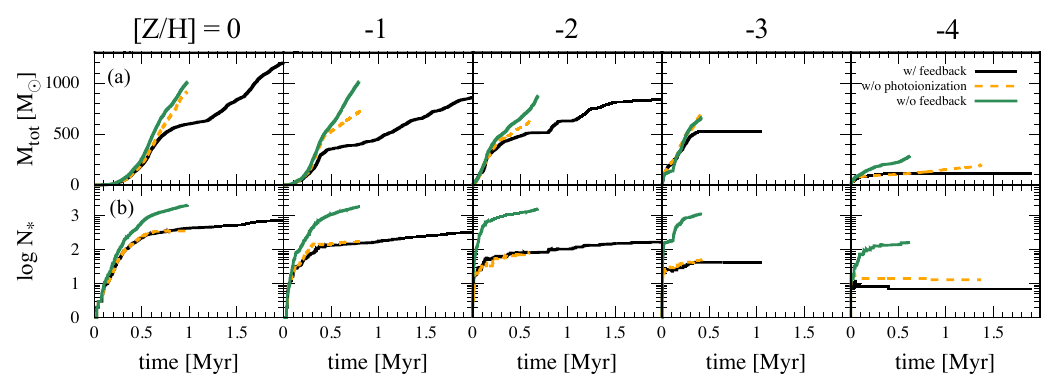}
\caption{
Metallicity dependence of the total mass and number of stars observed in our simulations as functions of the time, and effects of stellar radiative feedback on them. 
(a) Evolution of the total stellar mass. The black and green lines represent the mass evolution when we include and do not include any stellar feedback, respectively. The yellow dashed lines show the evolution when we take into account stellar feedback other than the ionizing feedback.
(b) The total number of stars. The colors of the lines have the same meaning as in panel (a). 
}
\label{fig::mass_evolution}
\end{figure*}

\begin{figure*}
 \includegraphics[width=0.9\textwidth]{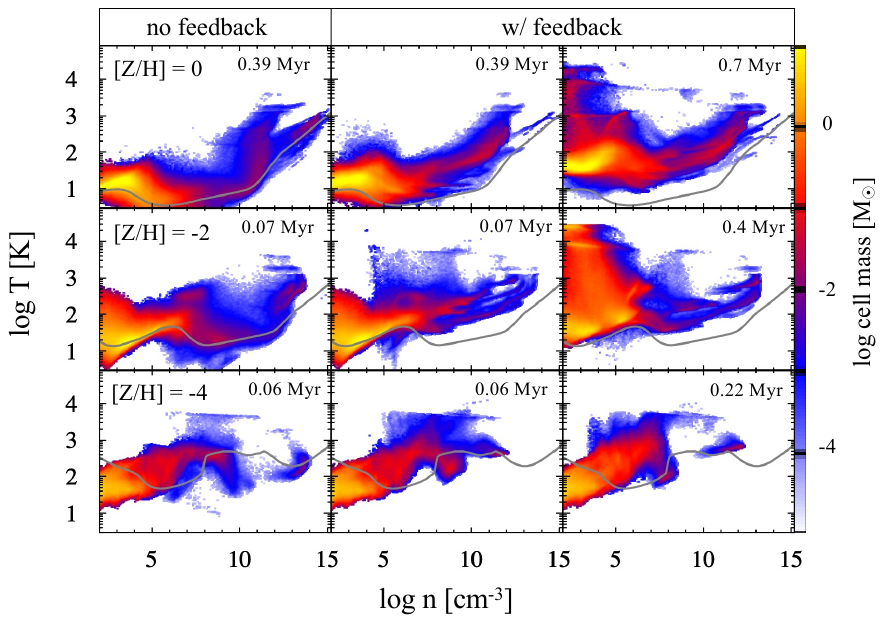}
\caption{
The phase diagram of the temperature versus density plane for [Z/H]$=0$ (top), $-2$ (middle), and $-4$ (bottom rows). 
The color represents the mass inside the cell, where we divide the shown region by $200\times200$ cells.
The grey solid lines show the temperature evolution of the collapsing cloud core estimated by one-zone calculation.
The left and middle panels illustrate the runs with no feedback and with feedback, respectively, at epoch A (see Figs.\ref{fig::snapshots_m0} -- \ref{fig::snapshots_m-4}).
In the runs with feedback at this epoch, the stellar mass reaches $100~M_\odot$ for [Z/H]$=-2$ and $0$ and $64~M_\odot$ for [Z/H]$=-4$. 
The right panels show the results of the runs with feedback at epoch B when the star formation rate starts to decrease.
}
\label{fig::rhoT_profiles}
\end{figure*}

\begin{figure*}
\includegraphics[width=0.75\textwidth]{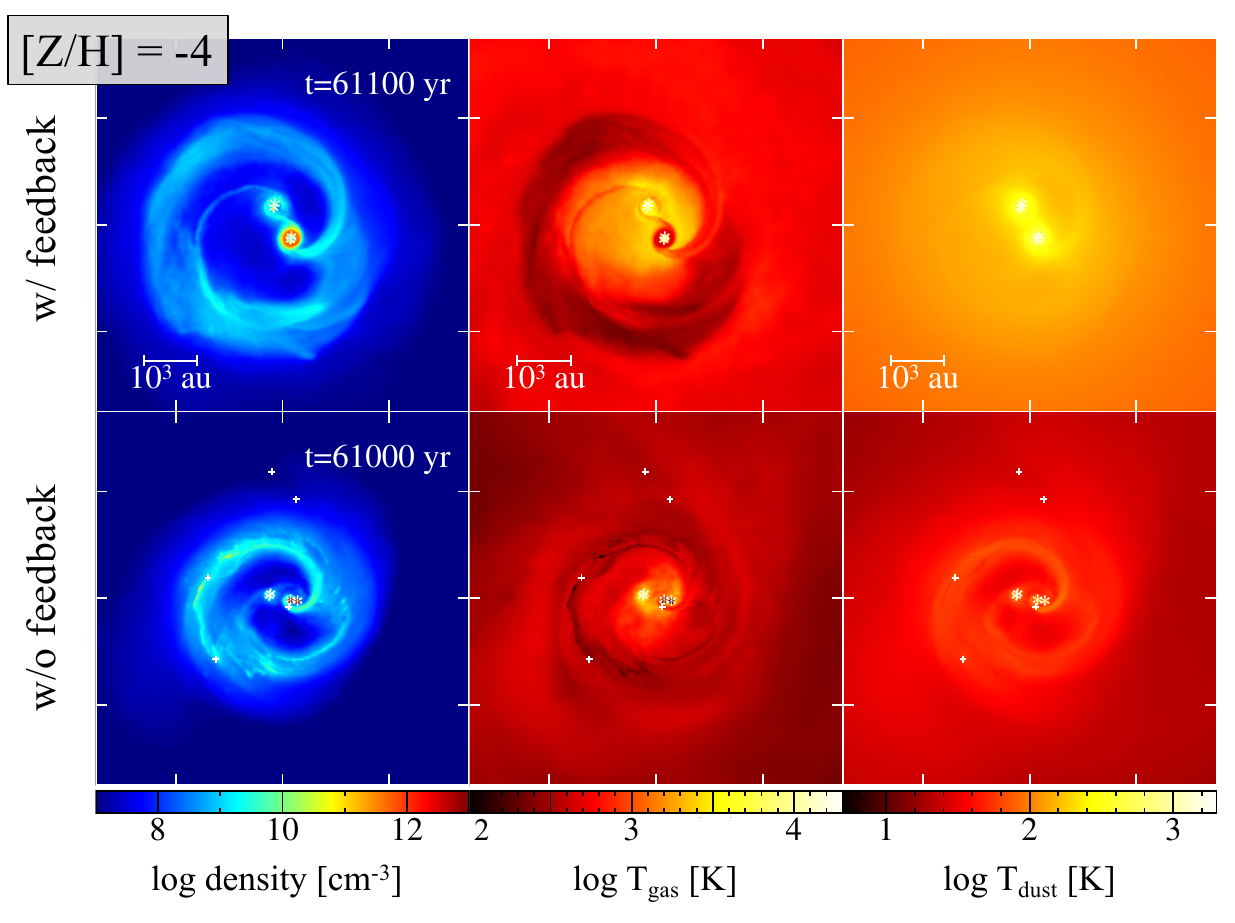}
\caption{
Effects of dust heating for the extremely metal-poor ([Z/H] = $-4$) case at the scales comparable to the protostellar disk surrounding the most massive stars. The upper and lower panels show the snapshots for the cases with and without radiative feedback, respectively, at the epoch of $t = 6.11 \times 10^4$ years after the first protostar formation. The left, middle, and right columns show projected distributions of the density, gas temperature, and dust temperature, respectively, for which the line of sight is orthogonal to the gas disc formed around the central massive stars.
White asterisks (dots) represent stars with masses higher (lower) than $1~M_\odot$.
}
\label{fig::dust_heating}
\end{figure*}

\subsection{Overview of the star cluster formation}

Fig.~\ref{fig::snapshots_coll} summarizes the overall evolution for all our simulations. The upper and middle rows show the density and temperature distributions when the first protostar is formed in each simulation run. In agreement with Paper I, the cloud morphology depends on metallicity in the following manner. When the metallicity is relatively high, at [Z/H] $\gtrsim -2$, turbulent motion creates a filamentary structure, which is efficiently cooled by metals \citep[e.g.][]{Li+2003, Jappsen+2005}. On the other hand, when the metallicity is lower, [Z/H] $\lesssim -3$, the gas temperature is higher due to inefficient cooling, and the increased gas pressure tends to erase the small structures created by turbulence, leaving one or two compact cores around the centre of the cloud.

The bottom panels of Fig.~\ref{fig::snapshots_coll} display the density distributions after approximately one to two million years from the formation of the first protostar. The protostars, represented by asterisks ($>1 M_{\sun}$) and dots ($<1 M_{\sun}$), form and congregate around the centre of the cloud. At higher metallicities, the stars are distributed in a cluster-like fashion, but they are less concentrated in the centre at lower metallicities. At the lowest metallicity, only a few stars are formed in isolation rather than in a cluster. In all cases, the expansion of HII regions created by massive stars eventually terminates star formation: the gas in the low-density HII regions around the central stellar system is expanding, indicating that the entire cloud is being disrupted.

~

We now take a closer look at the formation of star clusters for three different cases: [Z/H] $=0$ (Section~\ref{sec::Z0}), $-2$ (Section~\ref{sec::Z-2}), and $-4$ (Section~\ref{sec::Z-4}). In Section~\ref{sec::radiation_global_SF}, we compare the results of these cases in terms of the total stellar mass and the number of stars, as well as the impact of stellar radiative feedback on evolution.

\subsubsection{solar metallicity case ([Z/H] $=0$)}
\label{sec::Z0}

The top panel of Fig.~\ref{fig::snapshots_m0} displays the time evolution of the total stellar mass in the solar-metallicity case ([Z/H] $=0$). The bottom three rows illustrate the projected density, temperature, and H$_2$ fraction at four different points in time (A to D, marked with asterisks in the top panel). We have not included the degree of ionization, as the fully ionized region is almost the same as the area with a temperature of $\gtrsim 10^4~$K.

At epoch A ($0.39~$Myr), fragmentation of a filamentary cloud leads to the formation of protostars in alignment. No massive star has formed yet, and no HII region is visible. By epoch B (at $0.7~$Myr), the most massive protostar has grown to a mass of approximately $20$--$30~M_\odot$ and begins to emit copious ionizing photons, heating the surrounding regions to a temperature of a few thousand K. The hot gas disrupts the accretion flow onto massive stars, slowing their growth (see the change in slope of the time evolution shown in the top panel of Fig.~\ref{fig::snapshots_m0}). At the same time, streams of neutral gas heading toward the centre shield the outer regions from the stellar ionizing photons, thus preventing the expansion of the HII region.

At epoch C (at $1.2~$Myr), the stars are mainly concentrated in the central pc-scale region. The HII region expands in all directions, particularly in the lower-density areas that are perpendicular to the filament. The gas behind the filament is shielded from the ionizing radiation and remains neutral. By epoch D (at $1.7~$Myr), the entire domain is almost filled with the HII region, with some dense and cold gas still present. Although the gas near the surface of the filament is heated by radiation, the gas deep inside remains neutral or even fully molecular. The increased pressure from the heated gas compresses the neutral gas from the outside, leading to the continuation or resumption of star formation after epoch C (top panel, Fig.~\ref{fig::snapshots_m0}). The expansion of the HII region into the outer regions also promotes star formation thereby compression of clumps. As a result, the stellar population is spread out over a distance of several pc from the centre.

\subsubsection{low-metallicity case with [Z/H] $=-2$}
\label{sec::Z-2}

In this case, star formation generally proceeds similarly as in the [Z/H] $=0$ case, but with some differences; the expansion of the HII region occurs earlier and completely terminates star formation.
Fig.~\ref{fig::snapshots_m-2} illustrates the star formation history. Initially, no fragmentation is observed. As time progresses, however, star formation starts within a central pc region. By epoch B, the most massive stars have grown to $100~M_\odot$, and an HII region begins to expand. At epoch C, the entire simulated region is filled with the HII region, whereas star formation persists in the gas behind the filament. Subsequently, the gas near the filament surface evaporates due to photoheating, while the gas inside is compressed by the increasing ambient pressure. This triggers intermittent star formation that lasts until $t \sim 1.7~$ Myr. Star formation continues for an additional $\sim 1~$Myr after epoch C and nearly half of the stars are formed in this late stage in terms of mass.

At the early stages of epochs A and B, stars are densely packed in the central $\sim$ pc area. This high concentration causes more effective feedback from massive stars than in the solar-metallicity case and leads to star formation being completely stopped by $t \simeq 1.7~$Myr. The powerful feedback also affects the stellar distribution in the later stages. After the dispersal of the filamentary cloud, dense gases remain in clumps spread over several-pc regions (epochs C and D). Star formation in some of these clumps produces a stellar population that is as extended as the initial cloud and more sparsely distributed than in the solar-metallicity case.

\subsubsection{extremely metal-poor case with [Z/H] $=-4$}
\label{sec::Z-4}

The star formation sequence, in this case, is depicted in Fig.~\ref{fig::snapshots_m-4}. At epoch A ($t=0.06~$Myr), the gas is distributed in a highly compact core-like structure around the centre, rather than in a filamentary structure as seen in the higher-metallicity runs with [Z/H] $\gtrsim-2$. This is due to the higher temperature and pressure, which eliminates the fine structure created by the initial turbulence. Stars are born within a small cloud core of the central $10^{-2}$ -- $10^{-1}$pc region. Some stars are then ejected due to multi-body gravitational interactions, travelling as far as $\gtrsim 1$pc away from the centre. Stars in the dense central region quickly increase in mass, reaching $\sim20$ -- $30~M_\odot$ within the initial $0.1~$Myr. After this period, star formation is hindered by stellar feedback.
The development of an HII region is not the only factor that can lead to disruption of the cloud. Expansion of an H$_2$ photo-dissociation region (PDR) in the bipolar directions around the centre also plays a role. At a later stage of epoch C (at $t\sim 0.54~$Myr), the HII region begins to expand and disrupts the entire cloud. This is caused by the ejection of a massive star because of stochastic multi-body interactions among the central massive stars. This star then migrates to an outer low-density region, where the HII region can easily expand and suppress star formation in this area. We speculate that this ejection of massive stars is more likely in lower-metallicity cases, as stars tend to be clustered more closely together, compared to high-metallicity cases where stars are formed along filaments and their distribution is more spread out. Initial turbulence also contributes to the ejection of stars by inducing misalignment between the disc orientation and the orbital axis of the accreting gas. This causes the disc orientation to change over time, resulting in misalignment among the orbital axes of the member stars. This compact and misaligned configuration is unstable and leads to the ejection of massive stars through multi-body interactions \citep[e.g.][]{Wang+2019, Fujii+2022}.

\subsubsection{Evolution of total mass and number of stars across metallicities}
\label{sec::radiation_global_SF}

Here, we provide an overview of the effects of stellar feedback on star formation for all the cases considered. Figs.~\ref{fig::mass_evolution}(a) and (b) illustrate the growth of the total stellar mass and the number of stars over time. The green, yellow, and black lines represent, respectively, the cases without feedback, with LW radiation and grain heating but do not include ionizing radiation, 
and with all the feedback processes, i.e., photoionization, LW radiation, and heating of dust grains. In all cases, the feedback suppresses the total stellar mass and number of stars, particularly in lower-metallicity cases.

As is evident in Fig.~\ref{fig::mass_evolution}(a), star formation is inhibited in different ways between [Z/H] $\gtrsim -3$ (Sections 3.1.1 and 3.1.2) and $=-4$ (Section~\ref{sec::Z-4}). For [Z/H] $\gtrsim -3$, the suppression is mainly caused by ionizing radiation, as indicated by the deviation of the black lines from the green and yellow lines. In contrast, for [Z/H] $=-4$, the main factor in the suppression is LW radiation, which is evident from the fact that the total stellar masses in the cases with all the feedback (black) and without the ionization feedback (yellow) are almost the same until $0.7~$Myr and are lower than in the no-feedback case (green line). After $0.7~$Myr, the ionization feedback becomes effective and halts star formation, with the total stellar mass in the case without the ionization feedback (yellow) becoming slightly higher than that with feedback (black).

Fig.~\ref{fig::rhoT_profiles} illustrates the thermal state of the gas around forming clusters for [Z/H] $=0$, $-2$, and $-4$, with and without radiative feedback. At epoch B (rightmost column), when the feedback effect is most prominent, the ionized gas is seen as a hot component with temperatures of $\gtrsim 10^4~$K in the cases of [Z/H] $=0$ and $-2$. However, for [Z/H] $=-4$, the gas is divided into two components: a high-density ($n \gtrsim 10^{10}~\mathrm{cm^{-3}}$) component of LW-shielded accretion flows onto stars and a low-density ($\lesssim 10^8~\mathrm{cm^{-3}}$) component of low H$_2$-column-density gas in the envelope. Initially, the gas is distributed in both components, but by $t=0.22~$Myr, the increase of LW radiation field and the expansion of the H$_2$ PDR causes the gas to become clearly separated. Within the PDR, the gas cannot accumulate towards the centre due to inefficient cooling, instead expanding adiabatically to $\sim 10^{7}~\mathrm{cm^{-3}}$ and suppressing star formation. The LW feedback is only effective in the lowest metallicity cases [Z/H] $\lesssim -4$, where H$_2$ is the main coolant instead of the fine-structure line cooling at higher metallicity \citep[e.g.][]{Omukai2000, Bromm2002, Omukai+2008}. It is worth noting that the LW radiation reduces the H$_2$ cooling while leaving the fine-structure-line cooling unaffected. \cite{Susa+2014} also reported that star formation is terminated by LW feedback in the primordial case.

The reduction in the number of protostars (Fig.~\ref{fig::mass_evolution} b) can be attributed to local heating of dust grains, rather than ionization heating in all cases. Fig.~\ref{fig::rhoT_profiles} also displays the thermal states of the gas during an early stage (epoch A) for cases with (middle) and without (left) feedback. It is evident that the feedback only affects the temperature at high densities ($n \gtrsim 10^8 ~\mathrm{cm^{-3}}$), which corresponds to a dense gas near massive stars. Even though the total stellar mass remains unaffected at this stage, the feedback decreases the number of stars (Fig.~\ref{fig::mass_evolution}b). We will explore the impact of local heating in more detail in Section~\ref{sec::dust_heating} below.

\subsection{The effect of stellar radiation on the gas dynamics} 
\label{sec::radiation_feedback}

In what follows, we consider how radiation feedback affects star formation, focusing on the local effects on thermal evolution on scales comparable to protostellar disks.
In Section \ref{sec::dust_heating}, we illustrate how dust heating by stellar radiation significantly suppresses the formation of low-mass ($M_* \lesssim 1~M_\odot$) stars and reduces the number of stars. In Section \ref{sec::EUV_feedback}, we investigate how ionizing radiation regulates the growth of massive stars.

\subsubsection{Fragmentation suppression by dust heating} 
\label{sec::dust_heating}

We here show that heating of dust grains reduces the number of stars at any metallicity (see Fig.~\ref{fig::mass_evolution}b). We first consider the lowest-metallicity case of [Z/H] $=-4$, where a dense compact core appears at the cloud centre. Fig.~\ref{fig::dust_heating} shows the distribution of gas density (left), temperature (middle), and dust temperature (right panels) on scales comparable to the protostellar disk surrounding the most massive stars at $t=0.06~$Myr (epoch A), when the total stellar mass is $\simeq 63~M_\odot$ and the mass of the most massive star is $\simeq 30~M_\odot$. 
The top and bottom rows are for the runs with and without stellar feedback, respectively. Among the stellar feedback processes, only dust heating affects the thermal state of the disk gas at this early stage. Both the stellar ionizing and LW photons are still ineffective in modifying the gas dynamics (Fig.~\ref{fig::snapshots_m-4}). 
A massive gas disc of a few $10^3~$au forms around the massive binary system in the centre in both cases. 
While small density-inhomogeneities are smoothed out and the number of stars is reduced in the case with feedback, the overall disc structure shows no significant differences between the two cases. 
This implies that disc fragmentation is suppressed by heating of dust grains. We confirm the same effect more clearly in the right panels; the dust temperature exceeds several hundred K when the stellar radiation is included, while it stays below $100~$K otherwise. Note that the gas temperature also increases as the gas and dust are thermally coupled by collisions. As seen in Fig.~\ref{fig::rhoT_profiles}, dust heating reduces the amount of cold ($\lesssim 100~$K) gas from which low-mass stars primarily form (Paper~I). Therefore, dust heating strongly suppresses their formation.

\begin{figure*}
\includegraphics[width=0.85\textwidth]{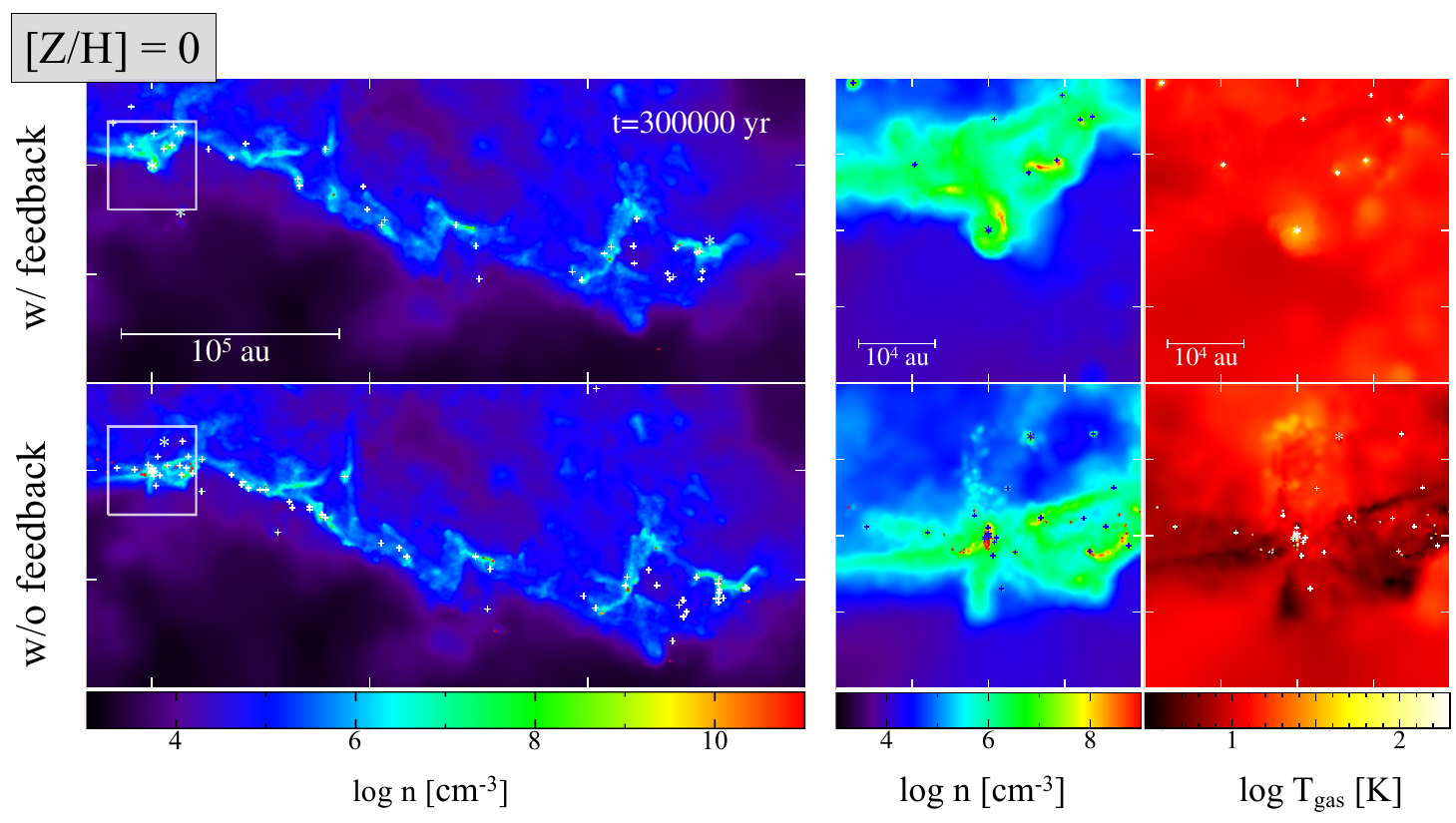}
\caption{
Effects of dust heating for the solar-metallicity ([Z/H] $=0$) case. 
The upper and lower panels illustrate the runs with stellar feedback and without feedback, respectively, at the epoch of $\simeq 0.3$~Myr after the first protostar formation (at epoch A in Fig.~\ref{fig::snapshots_m0}). The left columns show the projected density distributions on $\sim$pc scale. The middle and right columns show zoom-in views of the density and temperature distributions within the area delineated by the white square on the left panels. The asterisks (crosses) represent stars with masses higher (lower) than $1~M_\odot$.
}
\label{fig::filament_m0}
\end{figure*}

\begin{figure}
\centering
\includegraphics[width=0.4\textwidth]{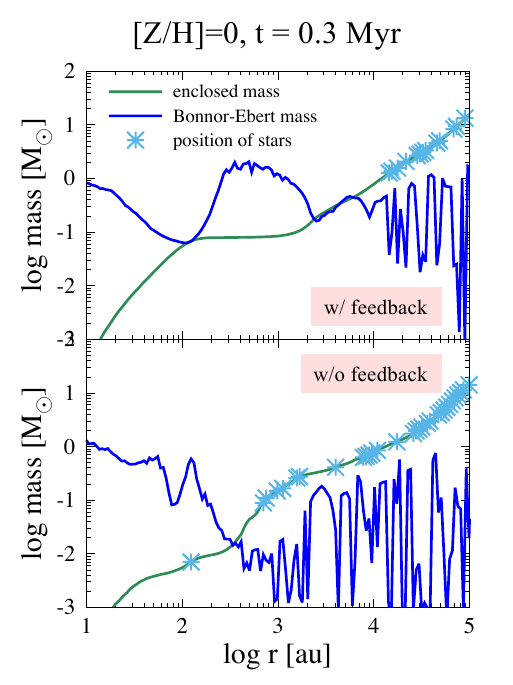}
\caption{
The enclosed mass (green) and the Bonnor-Ebert mass (blue lines) as a function of the distance $r$ from the most massive star for the solar metallicity [Z/H] $=0$ at $t=0.3~$Myr (i.e., epoch A in Fig.~\ref{fig::snapshots_m0}). 
The cases with and without the stellar feedback are shown in the top and bottom panels. 
The Bonnor-Ebert mass (equation~\ref{eq::BE}) is calculated by averaging the temperature and density inside the shell at the distance $r$.
The asterisks indicate the location of stars from the most massive star.
}
\label{fig::MBE_vs_Menc}
\end{figure}

\begin{figure}
\includegraphics[width=0.35\textwidth]{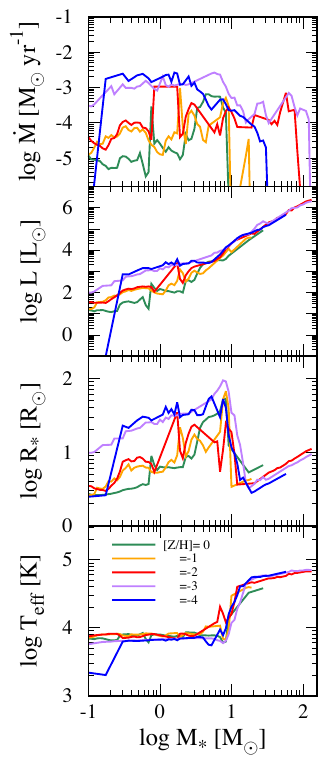}
\centering
\caption{
Evolution of the most massive star at the end of each run as a function of its mass.
From top to bottom, we show the mass accretion rate, stellar luminosity, radius, and effective temperature.
Different colors show different metallicity runs with [Z/H] $=0$ (green), $-1$ (yellow), $-2$ (red), $-3$ (purple), and $-4$ (blue).
}
\label{fig::stellar_props}
\end{figure}

\begin{figure*}
\includegraphics[width=0.95\textwidth]{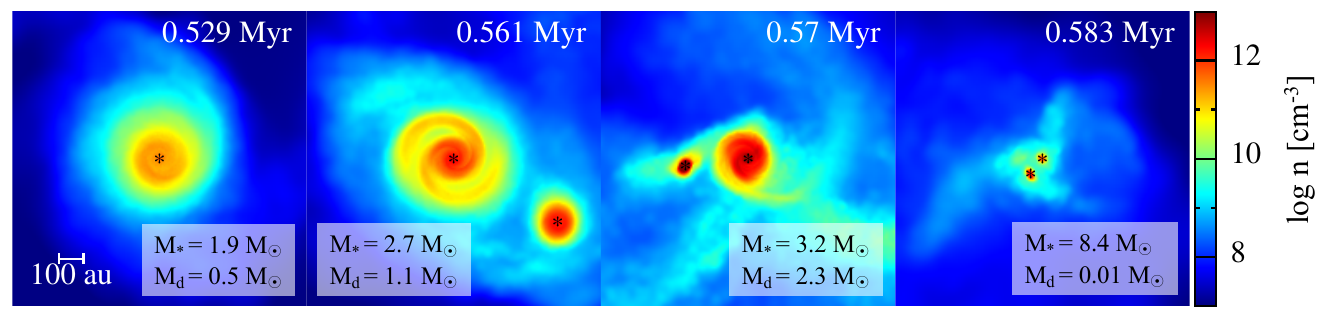}
\centering
\caption{
The face-on view of the projected density distribution around the most massive protostar at four different epochs for the solar metallicity run ([Z/H] $=0$). The asterisks show protostars with $M_* > 1~M_\odot$.
The masses of the central protostar ($M_*$) and its circumstellar disc ($M_\text{d}$) are given as legends.
}
\label{fig::disk_m0}
\end{figure*}

When the irradiation heating of dust grains is more efficient than the gas-dust collisional energy exchange, equation~\eqref{eq::Tdust} yields
\begin{align} 
T_\text{dust} \simeq T_\text{rad} &= \left ( \frac{L_*}{16\pi\sigma_\text{SB} r^2} \right )^{1/4} \nonumber \\ 
&= 320~\mathrm{K} \left ( \frac{L_*}{10^{38}~\mathrm{erg~s^{-1}}} \right )^{1/4} \left ( \frac{r}{10^3~\mathrm{au}} \right )^{-1/2} 
\label{eq::dust_heating}\\
&\simeq 380~\mathrm{K} \left ( \frac{M_*}{100~M_\odot} \right )^{1/4} \left ( \frac{r}{10^3~\mathrm{au}} \right )^{-1/2},
\end{align}
where $L_*$ is the stellar luminosity, and $r$ the distance from the star. 
The fiducial $L_*$ value $10^{38}~\mathrm{erg~s^{-1}}$ is for a $\simeq 30~M_\odot$ metal-free star (see Section~\ref{sec::EUV_feedback} below).
The last equation assumes the Eddington luminosity, which is valid for $M_* \gtrsim 100~M_\odot$.
In the absence of stellar irradiation, the gas temperature plummets from $500~$K to $100~$K at $\sim 10^{10}~\mathrm{cm^{-3}}$ due to dust cooling (Fig.~\ref{fig::rhoT_profiles}).
Equation~\eqref{eq::dust_heating} shows that radiative heating from a $30~M_\odot$ central star remarkably raises the temperature throughout the disc. The dust heating thus stabilises the disc and inhibits fragmentation induced by dust cooling.

Fig.~\ref{fig::filament_m0} illustrates a similar phenomenon for the solar-metallicity case ([Z/H] $= 0$), where a filamentary structure develops. 
The left panels show that dust heating has little effect on the pc-scale filamentary structures. It only affects the density structure within a radius of $\lesssim 10^4~$au from the source (see the white box in the left panels and its enlarged views in the right panels). By heating the gas and dust to several tens of K, stellar irradiation keeps the filament gravitationally stable and prevents fragmentation. Without irradiation, the temperature in the filament would decrease to several K by dust cooling, leading to fragmentation and low-mass star formation.

Fig.~\ref{fig::MBE_vs_Menc} shows the stabilising effect of dust heating by displaying the enclosed mass (in green) and the Bonnor-Ebert mass (in blue) as functions of the distance from the most massive star, $r$. The Bonnor-Ebert mass is the maximum amount of cloud mass that can be in a state of stable hydrostatic equilibrium for a given external pressure $p_0$:
\begin{align} \label{eq::BE}
M_\text{BE} = \frac{1.18c_\text{s}^4}{p_0^{1/2}G^{3/2}} = 0.12~M_\odot \left(\frac{T}{10~\mathrm{K}} \right )^{3/2} \left ( \frac{n}{10^6~\mathrm{cm^{-3}}} \right )^{-1/2},
\end{align}
where $c_\text{s}$ is the sound speed, $T$ the cloud temperature, and $n$ the number density. We have assumed a fully molecular gas with a mean molecular weight of $2.3$.
Fig.~\ref{fig::MBE_vs_Menc} shows that the temperature and thus $M_\text{BE}$ increase with stellar feedback. 
At distances $\lesssim 10^4~$au, $M_\text{enc} \lesssim M_\text{BE}$, or the cloud is gravitationally stable, and fragmentation does not occur at this scale. The position of the protostars (blue symbols) also indicates that fragmentation takes place only at $\gtrsim 10^4~$au.
Without stellar feedback, in contrast, $M_\text{enc}$ exceeds $M_\text{BE}$ at $\sim 10^3~$au, and, as a result, many protostars form outside this region.
Only one protostar is located at $100~$au where fragmentation is not expected. It is initially formed at $\gtrsim 10^3~$au away from the central star, but then migrates to $100~$au and forms a binary system (see Section~\ref{sec::binary}).

The characteristic mass of a protostar, $M_\text{ch}$, is determined by the fragmentation scale, which is the enclosed mass at the point of $M_\text{enc} = M_\text{BE}$. Fig.~\ref{fig::MBE_vs_Menc} shows that $M_\text{ch}$ increases from $0.01~M_\odot$ to $0.1$--$1~M_\odot$ when dust heating is present. This is in agreement with the analytic calculation of \citet{Sharda+2022}, who suggests that fragmentation at a scale below $M_\text{ch} \sim 1~M_\odot$ is inhibited at [Z/H] $=0$. They assumed a power-law density distribution around a protostar and calculated the temperature from the thermal balance between cooling and heating caused by stellar irradiation. As we will see in Section~\ref{sec::IMF}, $M_\text{ch}\sim 1~M_\odot$ indeed coincides with the peak in the mass distribution for [Z/H] $=0$.

\subsubsection{Evolution of massive stars under stellar UV feedback} 
\label{sec::EUV_feedback}

Fig.~\ref{fig::stellar_props} presents the evolution of the most massive star in each simulation run. It displays the mass accretion rate, luminosity, radius, and effective temperature as a function of the stellar mass, which is used as a proxy for time as it increases monotonically. Initially, the accretion rate is lower for higher metallicities, ranging from about $10^{-5}...10^{-4}~M_\odot~\mathrm{yr}^{-1}$ for [Z/H] $\gtrsim -2$ to $\sim 10^{-3} ~M_\odot~\mathrm{yr}^{-1}$ for [Z/H] $\lesssim 10^{-3}$. This is due to higher temperatures in environments with lower metallicity, as the mass supply rate from the accreting envelope $\dot{M}$ is related to the cloud temperature $T$ as 
\begin{align}
\dot{M} \sim 10\frac{c_\text{s}^3}{G}  = 1.4 \times 10^{-3}~M_\odot~\mathrm{yr^{-1}} 
\left( \frac{T}{100~\mathrm{K}} \right)^{3/2},
\end{align}
\citep{Whitworth&Summers1985},
where we adopt the factor of 10 in reference to \citet{Foster+1993}.
When [Z/H] $\gtrsim -2$, the accretion rate increases above $10^{-4}~M_\odot~\mathrm{yr^{-1}}$ for larger stellar masses. This is typical of higher-metallicity environments, where clouds break down into many protostars, and they competitively accrete the gas: a more massive star has a greater gravitational pull, leading to a larger Bondi radius and thus a higher accretion rate \citep[e.g.][]{Bonnell+2001}.

The stellar properties, apart from the accretion rate, are similar regardless of the metallicity, although there are some variations mainly in the pre-main sequence stage ($<10~M_\odot$) due to the difference in mass accretion rates. When the stars are less massive than $10~M_\odot$, they have large radii and low effective temperatures of a few thousand K. As the stellar mass approaches $10~M_\odot$, the stars experience a KH contraction and the effective temperatures increase to several $10^4 - 10^5$~K, leading to a burst of ionizing photons.

Accretion persists during the main-sequence stage for stars with low metallicities ([Z/H] $\lesssim -2$), while it ceases when the star reaches the main-sequence in higher metallicity cases ([Z/H] $\gtrsim -1$), as seen in the top panel of Fig.~\ref{fig::stellar_props}. This is mainly due to the different sizes of the circumstellar discs around massive stars. When [Z/H] $\lesssim -2$, the discs are relatively large and more shielded from photoionization feedback. As a result, mass accretion can continue even after the stellar mass exceeds tens of $M_\odot$.

At high metallicities of [Z/H] $\gtrsim -1$, the accretion rate suddenly drops around $10$ -- $20~M_\odot$. This is due to the depletion of the mass reservoirs in the parental cores, which is formed by the fragmentation of the large-scale filament. Once the accretion ceases, the protostar reaches the main-sequence stage, and the expanding HII region eventually halts the mass inflow, thus determining the final stellar mass.

Fig.~\ref{fig::disk_m0} shows that the solar metallicity run ([Z/H] $=0$) involves a binary system that is formed due to dynamical interaction with another core. This interaction causes non-axisymmetric perturbations on the disc, transferring angular momentum from it and leading to a high accretion rate of $10^{-3}~M_\odot~\mathrm{yr^{-1}}$. This causes the disc size to decrease from an initial $\sim 100~$au to below $10~$au. Initially, the disc gains mass from the environment, but then loses mass due to the binary interaction, reaching $M_{\rm d}=0.01~M_\odot$. At this stage, the protostar is still in the pre-main-sequence stage, and the ionizing photon emissivity is too low to disrupt mass accretion. When the accretion stops because of the binary interaction, the stellar KH contraction increases its UV emissivity, resulting in the expansion of the HII region and terminating the mass accretion. Subsequent mass growth is driven by stellar mergers, leading to a final stellar mass of $\simeq 30~M_\odot$.

\begin{figure*}
\includegraphics[width=1.\textwidth]{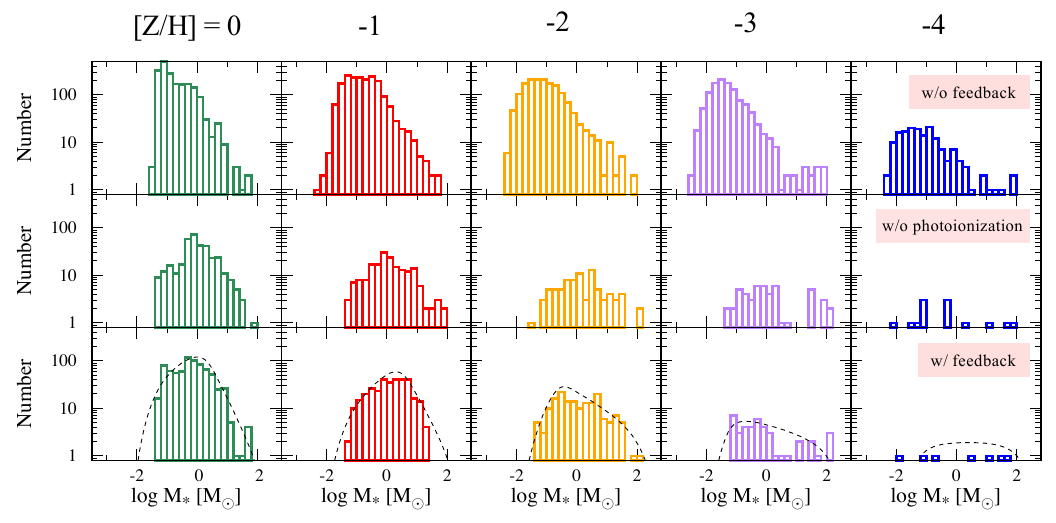}
\caption{
The stellar mass distributions at the end of our simulation for different metallicities with [Z/H] $=0$, $-1$, $-2$, $-3$, and $-4$ from left to right. The top, middle, and bottom rows show the results in runs without any stellar feedback, with stellar feedback but no photoionization feedback, and with all feedback effects. In the bottom panels, the dashed lines represent the fitted IMFs (equation~\ref{eq::IMF_tapered}) for the cases where stellar feedback is included.
}
\label{fig::mass_spectrum}
\end{figure*}

\begin{figure*}
\includegraphics[width=1.\textwidth]{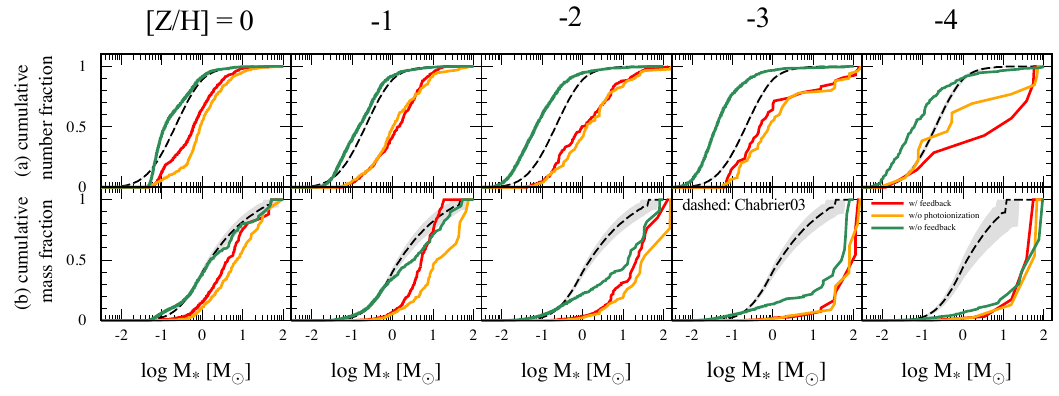}
\caption{
(a) The cumulative number fraction of stars with mass below $M_*$ to the total number of stars. The green lines show the cases with no feedback, while the red and yellow lines represent those with feedback, with the yellow lines illustrating the cases where no photoionization is considered. 
For comparison, we also show the values for the Chabrier IMF by the black dashed line.
The shaded regions in the data represent the 1-$\sigma$ error that occurs in sampling a finite stellar mass from the underlying Chabrier IMF. To assess this error, we created a stellar mass distribution by randomly sampling from the Chabrier IMF across $10^4$ realizations. The total stellar mass used for each sampling is based on the final stellar mass determined in our simulations with the same metallicity.
(b) Same as above, but showing the cumulative mass fraction in stars.
}
\label{fig::cumulative_mass_spectrum}
\end{figure*}

\begin{figure}
\includegraphics[width=0.47\textwidth]{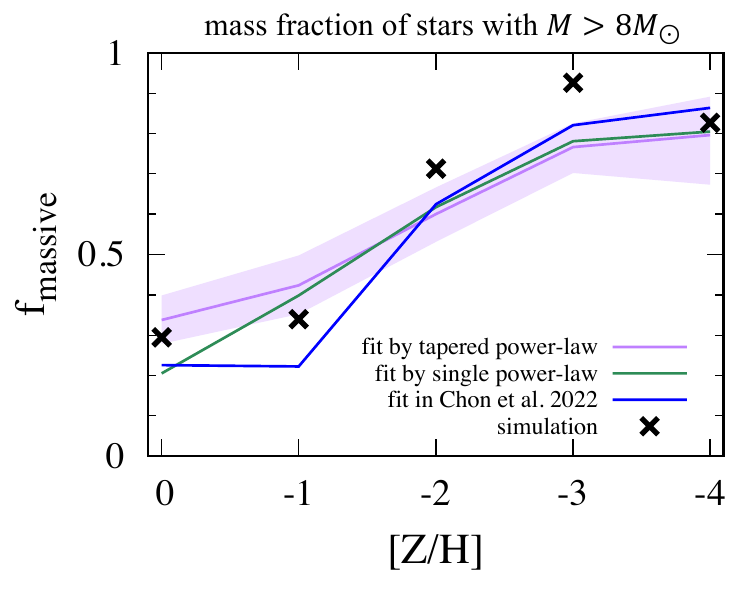}
\caption{
The mass fraction $f_\text{massive}$ in massive ($M_* > 8~M_\odot$) stars as a function of the metallicity [Z/H]. 
The crosses represent the results of our simulations.
The lines show those obtained 
by fitting the simulation results (with stellar feedback) with different functional forms: the single power-law (green line), ``tapered power-law'' (purple line), and the two-component IMF proposed in Paper II.
The blue line shows $f_\text{massive}$ from the fitted IMF in \citet{Chon+2022}, where the IMF is expressed with the composite of the Salpeter and log-flat components.
The shaded region indicates the 1 $\sigma$ error associated with sampling a finite stellar mass from the underlying mass distribution described by ``tapered power-law''. This error estimation is conducted similarly as in Fig. ~\ref{fig::cumulative_mass_spectrum}.
We note that the level of sampling errors in other fitted IMFs is comparable to the one described here.
}
\label{fig::f_massive}
\end{figure}

\subsection{IMFs at different metallicities} 
\label{sec::IMF}
\subsubsection{Overall shape of IMF}

Fig.~\ref{fig::mass_spectrum} shows the stellar mass distributions at the end of the simulations for all the metallicities. 
The bottom panels show the mass spectra when all the feedback processes are considered, where the simulation is followed until $t=2~$Myr ([Z/H]$=-1$ and $0$) or when the star formation is completely quenched due to feedback ([Z/H]$=-2$, $-3$, and $-4$).
For comparison, we show the spectra where we consider no feedback or the feedback without photoionization effect in top and bottom panels, respectively.
These comparison runs are stopped at $t\sim0.5$ -- $1~$Myr, at which the stellar feedback operates to decelerate the star formation in the runs that include all the feedback processes.
Fig.~\ref{fig::cumulative_mass_spectrum} shows the mass spectra in terms of the cumulative number (a) and the mass fraction (b). The dashed lines represent the Chabrier IMF \citep{Chabrier2003} in the solar neighbourhood for comparison.

By comparing the mass spectra in the top and middle rows of Fig.~\ref{fig::mass_spectrum}, it is evident that the population of low-mass stars with $M_* \lesssim 1~M_\odot$ decreases due to the dust heating (as discussed in Sec.~\ref{sec::dust_heating}). The fraction of low-mass stars (see Fig.~\ref{fig::cumulative_mass_spectrum} a) drops below 50\% at any metallicity when dust heating is included, while it is more than 90\% in number in cases without feedback. In the most metal-poor run with [Z/H] $= -4$, the mass spectrum becomes log-flat when dust heating is taken into account, similar to what is usually found for primordial star formation \citep[e.g.][]{Dopcke+2013, Susa2019, Chon+2021a}. 
In fact, when [Z/H] $=-4$ the only difference with respect to the primordial case is the effect of dust cooling at high densities $\gtrsim 10^{10}~\mathrm{cm^{-3}}$ (see Fig.~\ref{fig::rhoT_profiles}). 
In a dense region of the stellar neighbourhood, heating by stellar irradiation counteracts dust cooling as the stellar mass increases, thus suppressing fragmentation and low-mass star formation induced by dust cooling. As a result, the mass spectrum becomes very similar to the primordial one. When [Z/H] $\gtrsim -3$, the number of low-mass stars increases with metallicity. In these cases, the thermal evolution is largely different from the primordial one due to fine-structure line cooling of carbon and oxygen even under radiative feedback. As the metallicity increases further, the mass spectrum gradually changes from the log-flat distribution to a mass distribution that peaks at $\sim 1~M_\odot$.

Photoionization feedback has a significant effect on the number of massive stars, particularly at high metallicities [Z/H] $\gtrsim -2$. This is evident in the decrease of the mass fraction of stars with $M_* \gtrsim 1$--$10~M_\odot$ (Fig.~\ref{fig::cumulative_mass_spectrum} b). 
When [Z/H] $\lesssim -3$, the impact of ionizing radiation on the IMF shape is limited by the presence of a massive circumstellar disc (Fig.~\ref{fig::dust_heating}). This is because the HII region can only expand perpendicularly to the disc, allowing accretion to continue through the disc plane. On the other hand, when [Z/H] $\gtrsim -1$, stars tend to form in isolated small cores and are surrounded only by small discs, so the HII region easily expands and halts the accretion onto massive stars (Section~\ref{sec::EUV_feedback}). This reduces the number of massive stars and steepens the IMF slope at the high-mass end \citep[e.g.][]{He+2019}. At [Z/H] $=0$, ionization feedback modifies the high-mass slope from $-2$ to $-2.2$...$-2.3$, which is close to the slope observed in the solar neighbourhood (bottom panels in Fig.~\ref{fig::mass_spectrum}). Although the number of low-mass stars is smaller in our simulations than in the Chabrier IMF, the mass fraction of massive stars with $M_*\gtrsim 5~M_\odot$ is consistent with it (Fig.~\ref{fig::cumulative_mass_spectrum}b).

In summary, stellar feedback has a major effect on the IMF shape. 
Without feedback, the stellar IMF can be described by the combination of two components: a bottom-heavy Chabrier-like IMF and a top-heavy log-flat component, with the relative weight of the latter that depends on metallicity and redshift and that tends to zero when [Z/H] $\gtrsim -2$ and $z < 10$ (see also section~\ref{sec::f_massive}). 
With feedback, the IMF is best represented as a single power law at the high-mass end ($M_* \gtrsim 1~M_\odot$) with a cutoff at the low-mass end. As the metallicity decreases, the slope of the power law becomes shallower, and the IMF becomes more top-heavy.

We quantify the impact of metallicity on the IMF by fitting our IMFs with the analytic "tapered power-law" form with an exponential cutoff, as proposed by \citet{deMarchi+2005}:
\begin{align} \label{eq::IMF_tapered}
\phi(M_*) = \phi_0 M_*^{-\alpha} \left[ 1 - \exp \left(-\left( \frac{M_*}{m_0}\right)^c\right)\right] 
\exp\left(-\frac{m_\text{min}}{M_*}-\frac{M_*}{m_\text{max}}\right).
\end{align}
We fit the distributions by adjusting the three parameters, $\phi_0$, the overall normalization, $\alpha$, the slope at the high-mass end, and $m_0$, the peak mass, while we fix the other parameters $m_\text{min}=0.04~M_\odot$ and $m_\text{max}=150~M_\odot$, 
which contribute to the cutoff in the last exponential term at the low- and high-mass ends, respectively.
We set $c=1.6$, which determines the IMF slope at masses lower than $m_0$, in agreement with the original Chabrier value and that can reproduce the distributions for all the metallicities. We first calculate $\alpha$ by fitting the distribution at $M_* > 2~M_\odot$ with a power-law function $\phi(M_*) \propto M_*^{-\alpha}$. We then fit the IMF in the entire mass range with equation~\eqref{eq::IMF_tapered} and determine $m_0$ and $\phi_0$. Table~\ref{tab:IMF_fit} summarizes the best-fit values of $\alpha$ and $m_0$. The slope at the high-mass end gradually changes from a log-normal ($\alpha=1$) to a Salpeter-like value ($\alpha=2.23$).
The fitting functions derived for $\alpha$ and $m_0$ are
\begin{equation}
\alpha = 2.3 + 0.33 * [\text{Z/H}], \label{eq::alpha}
\end{equation}
\begin{equation}
\log_{10} m_0 = 0.2 + 0.45 * [\text{Z/H}], 
\label{eq::m0}
\end{equation}
which successfully represents the IMF variation from the primordial to present-day cases.
The metallicity dependence of the slope $\alpha$ is similar to that found by \citet{Guszejnov+2022}.
They performed cluster formation simulations resolving $\sim 10~$au scales in the metallicity range $-2 \lesssim \text{[Z/H]} \lesssim 0$
and found a power-law exponent $\alpha \simeq 2.3 + 0.3 * \text{[Z/H]}$ (equation 12 in their paper) by fitting the IMFs in the mass range $1\lesssim M_*/M_\odot \lesssim 10$ with power-law functions.

\subsubsection{Mass fraction in massive stars} \label{sec::f_massive}

In Fig.~\ref{fig::f_massive}, we quantify the effect of IMF variation on galaxy evolution by plotting the mass fraction of stars more massive than $8M_\odot$, $f_\text{massive}$. The black symbols represent the results of our simulations with stellar feedback, and the three lines indicate different fitting functions of the simulation data. The green and purple lines correspond to the fits assuming IMFs of the single power law with $\alpha$ (eq.~\ref{eq::alpha}) and `tapered power-law'' with $m_0$ (eq.~\ref{eq::IMF_tapered}) for the range in $0.3 < M_*/M_\odot < 100$, respectively. The blue line shows the fit given in Paper~II, which is a composite of the Salpeter ($\phi \propto M_*^{-2.3}$) for $0.3 < M_*/M_\odot < 100$ and the log-flat ($\phi \propto M_*^{-1}$) for $10 < M_*/M_\odot < 100$. This fit accurately reproduces the mass spectra when the feedback is not included. The fraction of the Salpeter component increases with metallicity, from zero at [Z/H] $\lesssim -4$ to unity at [Z/H] $ \gtrsim -1$.

The three different IMF fittings show a similar metallicity dependence of $f_\text{massive}$; $f_\text{massive}$ decreases from almost one at [Z/H] $=-4$ to 0.7 at [Z/H] $=-2$. 
At [Z/H] $=0$, it converges to the canonical value of $\sim 0.3$ found in the Salpeter IMF. The IMF fits for the results with and without stellar feedback give similar values for the fraction of massive stars. In fact, although the ionization feedback prevents the growth of massive stars and steepens the slope at the high-mass end, it only introduces a slight variation in terms of $f_\text{massive}$. The value at [Z/H] $=-1$ has a scatter among different fitting functions but within the range of the observed statistical scatter \citep{Bastian+2010}. We predict that the IMF at [Z/H] $=-1$ is slightly top-heavy, but further studies aimed at increasing the sample of simulated stars are needed to confirm this.

The peak mass of the IMF at $\text{[Z/H]}=0$ is around $1~M_\odot$, which is higher than the observed value $\simeq 0.2~M_\odot$ \citep{Chabrier2003}. This result is commonly found in RHD simulations without magnetic fields and outflows \citep[e.g.][]{Krumholz+2012}. Recent studies, however, have demonstrated that protostellar outflows can reduce stellar mass, with the IMF peak mass decreasing by a third while maintaining the same slope at the high-mass end \citep{Myers+2014, Cunningham+2018, Mathew+2021, Guszejnov+2021}. 
Including outflows would only affect the low-mass star fraction and leave the massive star fraction $f_\text{massive}$ unchanged. This can be seen from the comparison of the green and purple lines in Fig.~\ref{fig::f_massive}, where the purple line represents the case with an infinitely small turnover mass. This comparison shows that a smaller peak mass reduces $f_\text{massive}$ by $\lesssim 30\%$ at most when [Z/H] $=0$, while it has almost no effect on $f_\text{massive}$ when [Z/H] $\lesssim -1$.

The characteristic stellar mass $m_0$ decreases with decreasing metallicity, as seen in equation~(\ref{eq::m0}). This can be explained by smaller fragmentation mass in lower metallicity environments, which can be estimated from the Jeans mass at the density where dust cooling operates \citep{Omukai+2005, Schneider+2006}. Additionally, the IMF has a shallower slope at low metallicities than that of the Salpeter IMF in our simulation. This is in agreement with recent Gaia DR2 observations, which suggest that the mass distribution of low-metallicity halo stars ($-2.2 <$ [Fe/H] $ < -0.6$) can be fitted by a power-law distribution with $\alpha \sim 1.6$--$1.8$ in the mass range $0.1 \lesssim M_*/M_\odot \lesssim 1$ \citep{Hallakoun+2021}. A similar trend is also reported by \citet{Yasui+2023}, who studied a low-metallicity ([O/H] $= -0.5$) Galactic young cluster with $\sim 1500$ stars and derived $\alpha \simeq 2$ in the mass range $0.1 \lesssim M_*/M_\odot \lesssim 20$.

\begin{table}
    \centering
    \begin{tabular}{c||c|c|c|c|c}
        [Z/H] & 0 & -1 & -2 & -3 & -4  \\
        \hline
        $\alpha$ & 2.23 & 2.06 & 1.4 & 1.2 & 1.0 \\
        $m_0 / M_\odot$ & 1.46 & 2.3 & 0.2 & 0.07 & 0.01 \\
        SFE & 0.191 & 0.137 & 0.133 & 0.083 & 0.018 
    \end{tabular}
    \caption{The best fit IMF parameters in equation~\ref{eq::IMF_tapered} and SFE for different metallicities. $\alpha$: the slope at high-mass end, $m_0$: the turn-around mass.}
    \label{tab:IMF_fit}
\end{table}

\begin{figure}
\includegraphics[width=0.47\textwidth]{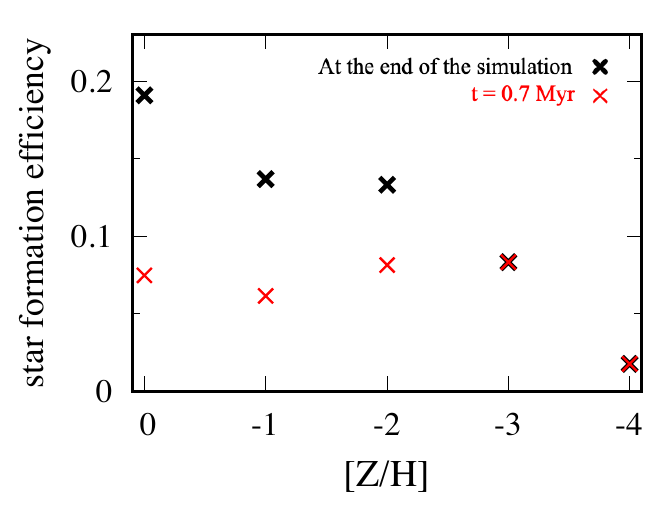}
\caption{
The star formation efficiency defined as $M_{*}/M_\text{cloud}$ as a function of metallicity [Z/H]. The black and red crosses indicate the star formation efficiencies at the end of the simulation and at $t=0.7~$Myr, when the feedback begins to operate, respectively.}
\label{fig::SFE}
\end{figure}

\begin{figure}
\includegraphics[width=0.47\textwidth]{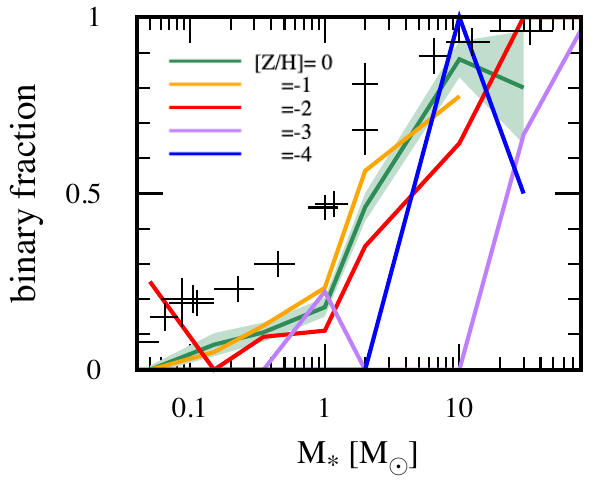}
\caption{
The binary fraction as a function of the stellar mass for [Z/H]$=0$ (green), $-1$ (yellow), $-2$ (red), $-3$ (purple), and $-4$ (blue lines). The black points show the observed value compiled by \citet{Offner+2022}. The shaded region indicates the 1-$\sigma$ error for [Z/H]$=0$.
}
\label{fig::binary_fraction}
\end{figure}

\begin{figure*}
\includegraphics[width=0.9\textwidth]{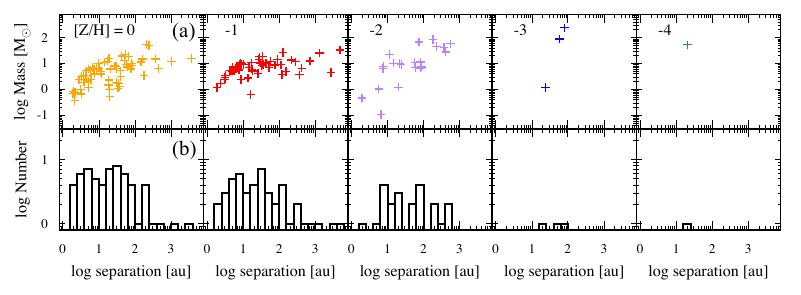}
\caption{
(a) The diagram of the binary separation versus the total stellar mass of the binaries for different metallicity cases.
(b) The distribution of the number as a function of the binary separation.
}
\label{fig::binary_m_r}
\end{figure*}

\subsection{Star formation efficiency at different metallicities} 
\label{sec::SFE}

Fig.~\ref{fig::SFE} shows the star formation efficiency (SFE), which defined by  $M_\text{*,total}/M_\text{cloud}$, where the total stellar mass $M_\text{*,total}$ is calculated at $t=0.7~\mathrm{Myr}$ (red) and at the end of the simulation (black). The epoch of $t=0.7~\mathrm{Myr}$ corresponds to the stage at which stellar feedback begins to decelerate star formation.
For the cases of $\text{[Z/H]} =-1$ and $=0$, the plotted values at the end of the simulations are lower limits, since star formation is still on going. It is evident that the SFE decreases as the metallicity decreases, in agreement with the findings of previous studies \citep{He+2019, Fukushima+2020b, Guszejnov+2022}.

Fig.~\ref{fig::SFE} shows that the SFEs at $t=0.7~\mathrm{Myr}$ are $\simeq 8\%$, almost independent of metallicity, except for the case of $\text{[Z/H]} = -4$. This suggests that the metallicity dependence of the SFE stems mainly from the later stages of evolution when the HII region expands and hinders the accumulation of gas. At $\text{[Z/H]}=-3$, star formation is quickly completed. However, when the metallicity is higher ($\text{[Z/H]} \gtrsim -2$), star formation continues and the SFE increases in a manner similar to ``triggered star formation by radiation'', where the higher pressure of the ionised gas compresses the clouds and accelerates star formation \citep[e.g.][]{Bisbas+2011}.

The metallicity dependence of the SFE is caused by three factors. First, due to less efficient cooling, the temperature of the HII region is higher at lower metallicity, making the cloud more vulnerable to disruption by expanding HII regions with higher pressure \citep[e.g.][]{He+2019, Fukushima+2020b}. This is illustrated in Fig.~\ref{fig::rhoT_profiles}, which shows that the temperature in the HII regions is $\lesssim 1$--$2\times 10^4~$K in $\text{[Z/H]} =0$, while it is $\sim 3\times 10^4~$K in $\text{[Z/H]}=-2$. Second, since the stellar mass distribution is more top-heavy at lower metallicity, stars emit more ionizing photons for a fixed cluster mass. Finally, small-scale dense structures do not form at low metallicities, and effective shielding of stellar ionizing photons does not occur. As a result, star formation can only continue for a shorter period of time, leading to a lower SFE in such environments.

Our SFE at [Z/H] $\lesssim -1$ is higher than those reported in earlier simulation studies. As the metallicity drops from $0$ to $-1$, the SFE decreases only by $20$--$30\%$ in our case, while \citet{He+2019} and \citet{Fukushima+2020b} show that it decreases by a factor of around $5$ and $3$, respectively. This discrepancy may be due to the higher spatial resolution of our simulations. By better-resolving structures such as dense gas discs and filaments, the shielding of stellar ionizing radiation may be more effective, resulting in more efficient star formation in our case.

\subsection{binary properties} \label{sec::binary}

Some stars in our simulations are in binaries or hierarchical multiple systems. If a pair of stars satisfies the following condition, we consider it to be a binary system \citep{Bate2012}:
\begin{align}
E_\text{grav} + E_\text{kin} < 0,
\end{align}
where $E_\text{grav}$ is the gravitational energy and $E_\text{kin}$ is the kinetic energy of the two stars. 
If two stars form a bound system with other stellar particles, we consider them to be part of a triple or higher multiple system. We define the binary fraction as that of stars in a given mass range that are part of binary, triplet, or quadruple systems.

Fig.~\ref{fig::binary_fraction} shows the binary fractions as a function of stellar mass for different metallicities. The black crosses represent the observational data in the solar neighborhood collected by \citet{Offner+2022}. Our results match the observed trend for [Z/H] $\gtrsim -2$, with the binary fraction gradually increasing with stellar mass and reaching unity around $M_* \gtrsim 10~M_\odot$. We also find little metallicity dependence for [Z/H] $\gtrsim -2$, which is consistent with previous simulations \citep{Bate2019, Guszejnov+2023}. When [Z/H] $\lesssim -3$, the distribution of the binary fraction is significantly different from those in higher-metallicity cases; almost no binary is found at $M_* \lesssim 1~M_\odot$, while the binary fraction is close to unity at $M_* \gtrsim 10~M_\odot$.

The difference in the mass-dependence of the binary fraction between [Z/H] $\gtrsim -2$ and $\lesssim -3$ is attributed to different origins of low-mass stars. When [Z/H] $\lesssim -3$, low-mass stars are formed through the fragmentation of a disc surrounding a massive star binary (or multiple). Consequently, low-mass binaries, if formed, are exposed to strong tidal forces (see Fig.~\ref{fig::dust_heating}) and are easily destroyed, with only tight binaries with separations smaller than the tidal radius being able to survive. Only massive stars are in hierarchical multiple systems around the disc centre. An exception is a decrease at $M_*\gtrsim 30~M_\odot$ for [Z/H] $=-4$, which is caused by incidental stellar ejection events through multi-body interactions.
On the other hand, when [Z/H] $\gtrsim -2$, stars are usually formed inside isolated cores created by filament fragmentation. The binary fraction increases with increasing stellar mass because massive stars are usually formed inside massive cloud cores, increasing the chance of forming binaries \citep{Bate2012}.

We note that at [Z/H] $\lesssim -3$ the binary fractions for $M_* \gtrsim 10~M_\odot$ are smaller than those found in Paper~I, where almost all massive stars are in binaries or multiple higher-order systems.
When [Z/H] $=-4$, for example, the binary fraction for $M_* \gtrsim 10~M_\odot$ decreases to 50\% from 100\% obtained in Paper~I. The reduction in the binary fraction stems from few-body interactions between massive stars. Since the simulations in this paper follow longer-term evolution than in Paper~I, there are more chances of stochastic interactions that eventually break binary or multiple systems.

Fig.~\ref{fig::binary_m_r} shows scatter plots of the total mass versus the binary separation (panel a) and histograms of the separation (panel b) for the binaries found in our simulations. At [Z/H] $=0$, the separation is distributed almost uniformly from approximately 1~au to 100~au. The decrease in number around 1~au is likely a numerical artifact due to the limited resolution of the sink radius, which is approximately 1~au. As the metallicity decreases, the typical separation increases, ranging from 10 to 1000~au at [Z/H] $=-2$. This is due to the larger fragmentation scales, or Jeans length, at lower metallicities, reflecting higher temperatures \citep{Inutsuka&Miyama1992}. However, observations show the opposite trend, with the fraction of close binaries (separation less than 10~au) increasing with decreasing metallicity \citep{Moe+2019}. One potential explanation for this discrepancy is that the typical initial conditions of star cluster formation may vary at low metallicities, leading to the formation of more close binaries. For example, higher surface densities of the initial clouds are more likely in low-metallicity environments, which increases the chance of close encounters and the formation of close binaries \citep{Guszejnov+2023}.

\section{Discussion} \label{sec::discussion}

\subsection{Comparison with previous studies}
We have observed a variation in the IMF, which shifts from Salpeter-like to more top-heavy as the metallicity decreases. We have identified a metallicity range of $Z \sim 0.01$--$0.1~Z_\odot$ below which the IMF is more top-heavy than the Salpeter type, which is consistent with our previous studies (Paper~I and II). Several studies have been conducted to understand the effect of stellar radiative feedback in low-metallicity environments. \citet{Matsukoba+2022} used hydrodynamical simulations to investigate the fragmentation of circumstellar discs at different metallicities, taking into account dust heating. They found that the average number of fragments increases with metallicity at [Z/H] $\lesssim -3$, with the average number of fragments being 10 in the primordial case and $12$--$18$ at [Z/H] $\lesssim -4$, although the metallicity dependence is less pronounced than in previous studies that did not consider dust heating \citep{Clark+2011, Dopcke+2011, Dopcke+2013}. \citet{Safranek-Shrader+2016} studied the mass spectrum at [Z/H] $=-2$, taking into account stellar irradiation of dust grains. They followed the evolution for an initial period of $18~$kyrs and concluded that the mass spectrum is given by a log-flat function with a cut-off at the high-mass end. This is in agreement with our result, although our cut-off mass scale is much higher since we followed the longer-term evolution of protostellar accretion.

Fragmentation of global filamentary structures is a critical factor in the variation of the IMF with metallicity. In the case of [Z/H] $=-2$, the boundary between a top-heavy and Salpeter IMF, filaments are stabilized by H$_2$ formation heating at a temperature of $50$--$100~$K (see Fig.~\ref{fig::rhoT_profiles}) and do not fragment into low-mass stars. The thermal physics used can influence the resultant IMF in this particular metallicity case. Other studies have predicted Salpeter-like IMFs at [Z/H] $=-2$ \citep{Bate2019, Tanvir+2023}. The discrepancy between our result and theirs may be due to the assumed initial H$_2$ abundance and the adopted prescription of the dust-gas thermal coupling. For example, \citet{Bate2019} started simulations from fully molecular clouds and found that filaments fragment effectively as the temperature is significantly reduced due to the lack of H$_2$ formation heating. \citet{Tanvir+2023} found a Salpeter-like IMF at [Z/H] $=-2$ in their simulation assuming that the gas and dust temperatures are the same. The filaments in their simulation have lower temperatures ($\sim 10~$K) than ours and fragment efficiently with a typical fragmentation mass lower than ours. In our simulations, however, gas and dust thermally couple only above $n\gtrsim 10^7$--$10^9~\mathrm{cm^{-3}}$, below which the gas has a higher temperature than the dust \citep{Omukai+2005, Safranek-Shrader+2016, Bate2019}. This emphasizes the importance of precise thermal modeling to accurately determine the IMFs in low-metallicity environments.

At solar metallicity ([Z/H] $=0$), our results are broadly consistent with the literature. Several authors showed that stellar feedback reduces the number of low-mass stars with $M_* \lesssim 0.1~M_\odot$, resulting in the peak mass of $\sim 0.1$--$1~M_\odot$ \citep{Offner+2009, Krumholz+2012, Mathew+2021}.
Early attempts to incorporate radiation feedback suffered from an ``overheating problem'', where dust heating is too effective and the median stellar mass becomes too large as star formation proceeds, as the feedback increases the cloud temperature \citep{Bate2009, Krumholz+2011}.
\citet{Krumholz+2012} found that turbulence in molecular clouds can help to solve the problem; turbulence creates small-scale structures that allow stars to form in an isolated manner, and dust heating only affects the temperature locally without influencing the formation of other stars. As a result, the peak mass of the IMF remains almost unchanged over time. In our simulations, the median mass at [Z/H] $=0$ is approximately 0.6 $M_\odot$ and does not vary much, similar to the findings of \citet{Krumholz+2012}. This median mass is three times larger than the Chabrier IMF \citep{Chabrier2003}. If protostellar outflows are taken into account, the typical mass will decrease, as in previous simulation studies \citep{Li+2010, Krumholz+2012, Guszejnov+2021}.

Our simulations have also shown the SFE variation with metallicity from the extremely metal-poor to the present-day universe. When the metallicity is as low as [Z/H] $=-4$, the mass accretion is first reduced due to the H$_2$-dissociating LW radiation \citep{Susa2013, Susa+2014} and then is terminated eventually by the ionizing radiative feedback when the total stellar mass reaches $\sim100~M_\odot$, similar to the case of primordial star formation \citep[e.g.][]{McKeeTan2008, Hosokawa+2011, Hirano+2014, Hirano+2015, Sugimura+2020, Latif+2022}. At [Z/H] $\gtrsim -3$, the LW radiation has less impact and the SFE increases by a factor of $\sim 5$ from the case of [Z/H] $=-4$. Our SFE of $\sim 10$--$20\%$ at [Z/H] $=0$ is consistent with those found in the literature \citep{Geen+2017, He+2019, Fukushima+2020b, Kim+2021, Guszejnov+2022}.

Our simulations lack several effects that play important roles in the present-day universe, including magnetic fields, protostellar outflows, and radiation pressure. Previous studies have demonstrated that magnetic fields can provide support to clouds, preventing them from collapsing due to gravity. The combined effect of the magnetic fields and protostellar outflows reduces the SFE by a factor of a few \citep{Hansen+2012, Myers+2014, Cunningham+2018}. The radiation pressure on dust grains can hinder the mass accretion, reducing the masses of individual massive stars by a factor of $0.6$ -- $0.7$ when [Z/H] $= -3$ and $-2$ \citep{Fukushima+2020a} although its effect on cluster-scale star formation is rather limited unless the surface density of the initial cloud is very large, exceeding $\sim 10^5~M_\odot~\text{pc}^{-2}$ \citep{Crocker+2018}.

\begin{figure}
\centering
\includegraphics[width=0.35\textwidth]{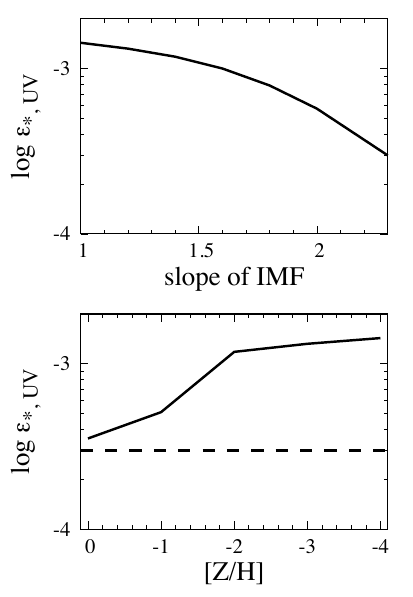}
\caption{
(a) The normalized UV emissivity ($\epsilon_{*,\text{UV}}$) as a function of the slope of the Salpeter IMF, $\alpha$. 
(b) The normalized UV emissivity as a function of the metallicity. We assume the metallicity-dependent IMF approximated by single power-law functions whose slopes are provided in Table~\ref{tab:IMF_fit}. The dashed horizontal line represents the emissivity for the canonical Salpeter IMF, for which $\alpha = 2.3$.
}
\label{fig::ML_ratio}
\end{figure}

\subsection{Implications for observations} \label{sec::JWST}

We have shown that the IMF becomes significantly top-heavy at [Z/H] $\lesssim -2$. To test this hypothesis, it may seem straightforward to study stars in dwarf galaxies in the Local Group, some of which have metallicity in this range \citep[e.g.][]{Kirby+2013}. This is, however, challenging since most of the massive stars in those galaxies have already ended their lives and only the mass distribution of low-mass stars, $M_* \lesssim 0.77~M_\odot$ \citep{Geha+2013}, can be inferred. Examining the IMF by resolving individual stars is possible only for nearby star-forming regions with metallicity above [Z/H] $\simeq -2$. For instance, \citet{Schneider+2018} derived an IMF with a slope $\alpha \simeq 2$ in the 30 Doradus Nebula in the Large Magellanic Cloud. In the outer Milky-Way young cluster Sh2-209 with low metallicity of [O/H] $=-0.5$, \citet{Yasui+2023} found a power-law distribution with $\alpha \simeq 2$ down to $M_* \gtrsim 0.1~M_\odot$. These observations imply that the IMF can be somewhat top-heavy in low-metallicity environments, which is in agreement with the results of our study. \citet{Marks+2012} discussed the IMF of stars in globular clusters by analyzing their radial distribution. They concluded that the IMF slope changes from $2.3$ at [Z/H] $=0$ to $1.3$ at [Z/H] $=-2$ at the high-mass end, consistent with our findings.

Recently, extremely metal-poor galaxies (EMPGs) with active star formation with metallicity [Z/H] $\sim -2$ to $-1$ have been discovered \citep{Izotov+2019,Kojima+2020}. Two of them have somewhat peculiar chemical abundance, [Fe/O] comparable to the solar abundance, which cannot be explained by the superposition of core-collapse SN yields. \citet{Isobe+2022} demonstrated that energetic explosions such as hypernovae or pair-instability SNe can provide the necessary abundance of Fe to match the observed ratio. This implies that very massive stars, which were the progenitors of such explosions, may have been formed
in the EMPGs in a top-heavy mass distribution.

\begin{figure}
\centering
\includegraphics[width=0.45\textwidth]{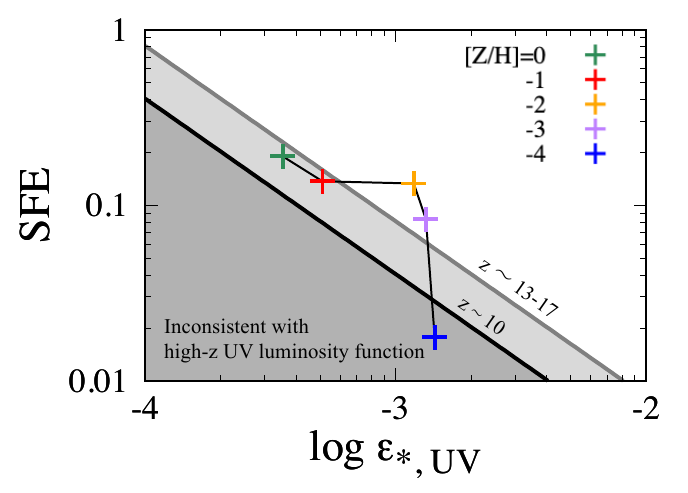}
\caption{
The scatter plot of the SFE and the normalized UV emissivity. The crosses represent values predicted by our simulations. The different colors represent different metallicities of [Z/H] $=0$ (green), $-1$ (red), $-2$ (yellow), $-3$ (purple), and $-4$ (blue). The shaded region below the black (grey) line denotes the parameter space where the expected UV luminosity function at $z\sim10$ ($z\sim 13$--$17$) is inconsistent with photometric observations by JWST \citep{Inayoshi+2022}.
}
\label{fig::epsilon_to_SFE}
\end{figure}

Recent JWST observations have indicated that galaxies at $z \gtrsim 10$ may have top-heavy IMFs. This is evidenced by the number density of bright galaxies in those epochs, which is higher than predicted by standard galaxy formation models \citep{Finkelstein+2022, Donnan+2023, Harikane+2023, Harikane+2023b, Furtak+2023}. This discrepancy may be due to a missing physical process in galaxy formation, such as metallicity-dependent IMFs \citep{Kannan+2022, Yajima+2022, Trinca+2023}. Our simulations have shown that top-heavy IMFs are likely to be realized at $-2 \lesssim$ [Z/H] $\lesssim -1$, which is the typical metallicity of the observed galaxies \citep{Nakajima+2023, Curtis-Lake+2023}. These IMFs increase the specific UV emissivity per stellar mass, thus increasing the number of UV-luminous galaxies \citep{Inayoshi+2022}.
To quantify how the UV emissivity changes with the IMF, we define the 
UV-luminosity conversion efficiency normalized by star formation rate (SFR) as
\begin{align}
\epsilon_{*,\text{UV}} \equiv \frac{L_\text{UV}}{\text{SFR} c^2} = 1.75 \times 10^{-4} \left ( \frac {L_\text{UV}}{10^{43} \mathrm{erg s^{-1}}} \right) \left ( \frac{\text{SFR}}{M_\odot \mathrm{yr}^{-1}} \right )^{-1},
\end{align}
where $L_\text{UV} \equiv \nu_0 L_{\nu_0}$ and $L_{\nu_0}$ the specific luminosity in units of erg s$^{-1}$ Hz$^{-1}$ at the frequency $\nu_0$ corresponding to the wavelength $1500~$\AA~\citep{Madau&Dickinson2014}. 
We assess the effect of the IMF on the UV emissivity, $\epsilon_{*,\text{UV}}$, by assuming a constant star formation rate and using the stellar population synthesis code, {\tt STARBURST99} \citep{Leitherer+1999}. We follow the evolution for $10~$Myr using the stellar evolutionary tracks for $0.1~Z_\odot$, which is the typical value suggested for JWST galaxies \citep{Harikane+2023}. We approximate the IMF with a single power-law $\phi \propto M_*^{-\alpha}$ in the range of $0.3~M_\odot < M_* < 100~M_\odot$. Fig.~\ref{fig::ML_ratio}(a) displays the normalized UV emissivity $\epsilon_{*,\text{UV}}$ as a function of the IMF slope, $\alpha$. We observe that $\epsilon_{*,\text{UV}}$ increases as the IMF slope decreases, meaning that $\epsilon_{*,\text{UV}}$ is larger for more top-heavy IMF. The log-flat IMF yields the UV emissivity 
four times higher than the Salpeter IMF, in agreement with previous studies \citep{Inayoshi+2022, Harikane+2023}.
Fig.~\ref{fig::ML_ratio}(b) illustrates the variation of UV emissivity with metallicity, for which we use single power-law IMFs with the exponents given in Table~\ref{tab:IMF_fit}. A sudden change in $\epsilon_{*,\text{UV}}$ is visible between [Z/H] $=-1$ and $-2$. For [Z/H] $\gtrsim -1$, $\epsilon_{*,\text{UV}}$ is close to the Salpeter-IMF value, while it is three or four times higher for [Z/H] $\lesssim -2$. This implies that the UV emissivity of high-redshift galaxies observed by the JWST, which typically have metallicities of [Z/H] $\sim -1$ or $-2$, can be significantly increased due to the top-heavy nature of the IMFs.

Fig.~\ref{fig::epsilon_to_SFE} shows our simulation results on the SFE-$\epsilon_{*,\text{UV}}$ plane. The cross symbols represent our simulation results, and the shaded region below the black (grey) line indicates the region where the pairs of SFE and $\epsilon_{*,\text{UV}}$ fall short of the number of JWST luminous galaxies at $z\sim 10$ ($z\sim 13$--$17$, respectively), taken from \citet{Inayoshi+2022}. We find that the combination of SFE and $\epsilon_{*,\text{UV}}$ in our simulations can explain the high number density of luminous galaxies in the metallicity range $-3 \lesssim \text{[Z/H]} \lesssim -1$. This range is consistent with the observed values of the JWST galaxies \citep[e.g.][]{Nakajima+2023}. At higher metallicity of [Z/H] $\gtrsim -1$, our UV emissivity is smaller than required because the IMF is less top-heavy. Conversely, at lower metallicity of [Z/H] $\sim -4$, the small SFE provides only a small stellar content in a given halo, leading to insufficient number densities of the luminous galaxies.  It should be noted that our SFE should be considered an upper limit, as \citet{Inayoshi+2022} defined the SFE as the stellar conversion efficiency from the total halo gas, a small part of which will become dense star-forming clouds assumed as our initial conditions. To accurately estimate the SFE, further studies are necessary to take into account the galaxy-scale gas dynamics and star formation activity in the early universe.

The UV emissivity increases further for [Z/H]$\gtrsim -1$ at $z \gtrsim 10$ if we take into account the additional heating effect caused by the CMB, as discussed in previous studies \citep{Schneider&Omukai2010, Chon+2022}. In the early universe, the CMB temperature rises to $ 2.76~(1+z)$K, establishing a minimum temperature for dust grains. This prevents the gas from cooling below the CMB temperature, thereby stabilizing the gas against fragmentation. Consequently, in the early universe, typical mass scales for fragmentation become larger \citep{Schneider&Omukai2010}.

In our previous work (Paper II), we have demonstrated that at $z \gtrsim 10$ the CMB radiation can suppress fragmentation and a top-heavy IMF is realized even at [Z/H] $= -1$, where the IMF would otherwise follow the present-day-like form without the CMB radiation. In such cases, the IMF can be decomposed into the Salpeter-like component and the log-flat top-heavy component. At $z=10$ ($z=15$), the top-heavy component dominates $40\%$ ($60\%$) of the total stellar mass, as detailed in equation 10 of Paper II. This doubles the mass-to-light ratio or $\epsilon_{*, \text{UV}} \sim 8\times 10^{-4}$ ($10^{-3}$) for $z=10$ ($z=15$, respectively).
Figure~\ref{fig::epsilon_to_SFE} illustrates how such an increase in UV emissivity makes the luminosity function at [Z/H]$=-1$ consistent with the observed number density of UV-luminous galaxies, both at $z=10$ and around $z \sim 13$ -- $17$ \citep{Trinca+2023}. In our upcoming paper, we will explore how CMB heating affects the IMF and SFE during the era of JWST galaxies.

\section{Summary} \label{sec::summary}

We have studied the formation of star clusters in environments with different metallicities using radiation hydrodynamics simulations. 
In particular, we have followed the long-term evolution for two million years after the first protostar formation, during which stellar radiative feedback effects become significant. 
We have considered the stellar mass spectrum at the end of each simulation run as the stellar IMF. The derived IMFs are approximated by a single power-law function with truncation at the low-mass end, whose shape varies with different metallicities. The IMF gradually varies from the log-flat distribution at [Z/H] $=-4$ to a Salpeter-like form above a metallicity threshold between [Z/H] $=-2$ and $-1$. The mass fraction that contributes to supernovae ($M_* > 8~M_\odot$) decreases from $70\%$ at [Z/H]$=-2$ to $30\%$ at [Z/H]$=-1$.

Stellar feedback has a significant impact on the formation of both low-mass and high-mass stars. Heating of dust grains by stellar irradiation increases the temperature of the gas in the vicinity of massive stars, suppressing fragmentation and reducing the number of low-mass stars. Additionally, ionizing radiation from high-mass stars creates HII regions and hinders stellar growth, thus decreasing the number of high-mass stars. The combination of those effects leads to a sharp peak in the IMF, particularly when [Z/H] $\gtrsim -2$. When [Z/H] $\lesssim -2$, feedback caused by the expansion of HII regions is weakened, as a massive circumstellar disc blocks the stellar ionizing radiation, resulting in an IMF that extends up to $\gtrsim 100~M_\odot$.

Our findings demonstrate that stellar radiative feedback is the primary regulator of star formation, particularly photoionization feedback caused by the expansion of HII regions for [Z/H] $\gtrsim -3$. The SFE decreases with decreasing metallicity due to three effects: higher temperature and pressure of HII regions at lower metallicities, a more top-heavy IMF which increases the stellar UV emissivity per unit mass, and a smoother neutral gas density distribution due to less efficient radiative cooling. The case of [Z/H] $=-4$ is unique in the sense that star formation is mainly suppressed by H$_2$-dissociating radiation, while photoionization feedback is effective only in completely terminating star formation. This results in a top-heavy, log-flat IMF similar to those proposed for primordial star formation \citep[e.g.][]{Chon+2021a}.

Our findings provide an explanation for the discrepancy between JWST observations and standard galaxy formation models: a higher number of UV luminous galaxies at $z \gtrsim 10$ than expected by theoretical models \citep[e.g.][]{Harikane+2023}. We suggest that a top-heavy IMF may provide a solution for resolving this tension, as it leads to a higher UV emissivity. Our results show that UV emissivities and SFEs in the range of $-3 \lesssim$ [Z/H] $\lesssim -2$ are sufficient to match the number of luminous galaxies observed by JWST. Interestingly, these metallicities are consistent with those observed for the most distant JWST galaxies.

%%%%%%%%%%%%%%%%%%%%%%%%
%%%%%%%%%%%%%%%%%%%%%%%%
\section*{Acknowledgements}
We thank Kazuyki Sugimura, Gen Chiaki, Hajime Fukushima, Steven Finkelstein, and Kaitlin Kratter for fruitful discussions and comments.
This work is financially supported by
the Grants-in-Aid for Basic Research by the Ministry of Education, Science and Culture of Japan 
(TH:19H01934, 19KK0353, 21H00041, KO:22H00149). 
RS acknowledges support from the Amaldi Research Center funded by the MIUR program Dipartimento di Eccellenza (CUP:B81I18001170001) and funding from the INFN TEONGRAV specific initiative.
We conduct numerical simulations on XC50 \texttt{ATERUI II} at the Center for Computational Astrophysics (CfCA) of the National Astronomical Observatory of Japan and XC40, through the courtesy of Prof. E. Kokubo.
We also carry out calculations on XC40 at YITP, Kyoto University.
We use the SPH visualization tool SPLASH \citep{SPLASH} in Figs.~\ref{fig::snapshots_coll}, \ref{fig::snapshots_m0}, \ref{fig::snapshots_m-2}, \ref{fig::snapshots_m-4}, \ref{fig::dust_heating}, \ref{fig::filament_m0}, and \ref{fig::disk_m0}.

\section*{DATA AVAILABILITY}
The data underlying this article will be shared on reasonable request to the corresponding author.

\bibliographystyle{mnras}
\bibliography{biblio2}

% Don't change these lines
\bsp	% typesetting comment
\label{lastpage}
\end{document}